\documentclass{emulateapj}
\usepackage[]{longtable}
\usepackage[]{natbib}

\newcommand{\f}{A\,0535+26\,}

\newcommand{\rxte}{{\it RXTE} }
\newcommand{\be}{\begin{equation}}
\newcommand{\ee}{\end{equation}}
\newcommand{\bdm}{\begin{displaymath}}
\newcommand{\edm}{\end{displaymath}}

\shorttitle{X-ray/optical observations of  A\,0535+26}
\shortauthors{Camero-Arranz et al.}

\begin{document}
\title{X-RAY AND OPTICAL OBSERVATIONS OF A\,0535+26}

\author{A. Camero-Arranz\altaffilmark{1,2}, M.H. Finger\altaffilmark{1,2}, C.A. Wilson--Hodge\altaffilmark{2}, P. Jenke\altaffilmark{2} ,  I. Steele\altaffilmark{5}, M.J. Coe\altaffilmark{6}, J. Gutierrez-Soto\altaffilmark{7,14}, P. Kretschmar\altaffilmark{8}, I. Caballero\altaffilmark{9},  J. Yan\altaffilmark{10}, J.Rodr\'{i}guez\altaffilmark{9}, J. Suso\altaffilmark{11}, G. Case\altaffilmark{12}, M.L.Cherry\altaffilmark{12}, S. Guiriec\altaffilmark{13} and V.A.McBride\altaffilmark{6}}

\affil{$^{1}$ Universities Space Research Association,  Huntsville, AL 35806\\
         $^{2}$  Space Science Office, VP62, NASA/Marshall Space Flight Center, Hunstville, AL 35812\\
         $^{5}$ Liverpool J. Moore's University\\         
         $^{6}$ University of  Southampton\\
         $^{7}$  Instituto de Astrof\'{i}sica de Andaluc\'{i}a, Spain\\
         $^{8}$ ESA/ESAC, Madrid, Spain\\         
         $^{9}$  AIM-CEA Saclay, Paris, France\\
         $^{10}$ Purple Mountain Observatory, Chinese Academy of Sciences, Beijing, China\\ 
         $^{11}$ University of Valencia, Spain\\         
         $^{12}$ Louisiana State University, USA\\
         $^{13}$  NASA GSFC, Greenbelt, MA\\
         $^{14}$ Valencian International University, 12006 Castell\'{o}n de la Plana, Spain}

\begin{abstract}

We present recent contemporaneous  X--ray and optical observations of the Be/X-ray binary system A\,0535+26 with the  \textit{Fermi}/Gamma-ray Burst Monitor (GBM) and several ground-based observatories.  These new observations are put into the context of the rich historical data (since $\sim$1978) and discussed in terms of the neutron star Be-disk interaction.  The Be circumstellar disk was exceptionally large just before the 2009  December giant outburst, which may explain the origin of the unusual recent X--ray activity of this source. We found a peculiar evolution of the pulse profile during this giant outburst, with the two main components evolving in opposite ways with energy.  A hard 30--70 mHz  X--ray QPO was detected  with GBM  during this 2009 December giant outburst. It becomes stronger with increasing energy and disappears at energies below 25\,keV.  In the long-term a strong optical/X--ray correlation was found for this system, however in the medium-term  the H$_\alpha$ EW and the V-band brightness showed an anti-correlation after $\sim$2002 Agust.  Each giant X-ray outburst occurred during a decline phase of the optical brightness, while  the H$_\alpha$ showed a strong emission.  In late 2010 and  before the 2011 February outburst,  rapid V/R variations are observed in the strength of the two peaks of the H$_\alpha$ line.  These had a period of $\sim$\,25 days and  we suggest the presence of a global one-armed oscillation to explain this scenario.  A general pattern might be inferred, where the disk becomes weaker and shows V/R variability beginning $\sim$\,6 months following a giant outburst.

\end{abstract}

\keywords{accretion,\,accretion disks\,---\,pulsars: individual
          (\f)\,---\,\,stars: emission-line, Be\,---\,\,stars:\,neutron\,---\,X--rays:\,binaries}

\section{INTRODUCTION}

X--ray binaries are composed of a donor star, usually still on the main sequence, and an accreting compact object either a neutron star or black hole. If periodic pulsations are detected from an X--ray binary, the compact object must be a neutron star and the system is called an accreting X--ray pulsar.  The largest sub-group of high mass X--ray binaries are the so-called  Be/X--ray binaries (BeXRB) in which the companion is a dwarf, subgiant or giant OBe star.  Be stars are rapidly rotating objects with a quasi-Keplerian disk around their equator. The ultimate cause of the formation of the disk is still under  debate  \citep{townsend04}. In  BeXRB  the optical and infrared  emission is dominated by Be star companion and characterized by spectral  lines in emission (particularly those of the Balmer series) and an IR excess,  revealing the physical state of the mass donor component.  In particular the detection of emission line features from the companion will confirm the presence of a disk around its equator \citep{coe06}. 

Most Be/X--ray binaries are transient systems. Historically, their outbursts have been divided into two classes. In the long-term, the X--ray variability of the transient BeXRB is characterized  by  type I (or normal) outbursts. These are regular and (quasi)periodic outbursts, normally peaking at or close to periastron passage of the neutron star, and reaching  peak luminosities L$_X\leq$ 10$^{37}$\,erg s$^{-1}$.  Type II (or giant) outbursts reach luminosities of the order of the Eddington luminosity for a neutron star (i.e., when the gravitational attraction balances the outward radiation force on the accreting material; L$_X\sim$10$^{38}$\,erg s$^{-1}$) \citep{frank02}  and become the brightest objects of the X--ray sky. During type II outbursts, an accretion disk may form. Unlike  normal outbursts,  giant outbursts have no consistently preferred orbital phase \citep{wilson08}. 

A\,0535+26 is a transient Be/X--ray binary pulsar, discovered by Ariel\,V  in 1975 (\cite{rosenberg75}, \cite{coe75}). The orbital period is  $\sim$111 days  with a pulse period  of $\sim$103\, s in a eccentric orbit (e$\sim$0.47, \citet[and references therein]{finger96}). The optical counterpart is the O9.7 IIIe star HDE\,245770 \citep{giangrande80}.  An exhaustive review of observations at different wavelengths from 1970 until 1989 is found in \citet{giovannelligraziati92}. 

This system shows both giant and normal outbursts.  The model of \citet{okazakinegueruela01} attempts to explain the origin of type I and II outbursts by the truncation of the circumstellar disk at a resonance radius between the disk keplerian velocity and the orbit of the neutron star.  Whether or not type I or II outbursts are produced depends on the eccentricity of the system.  For the case of the moderate eccentricity of \f (e=0.47) both types of behavior are possible as the disk expands and contracts between different resonant radii \citep{Haigh04, coe06}.  We would therefore expect some correlation between the disk status as inferred from $H\alpha$ equivalent width and profile and the presence or otherwise of X-ray outbursts.

High-dispersion optical spectroscopic observations of this source during and after the 2009 December giant outburst, showed drastic variabilities and indicated the existence of a warped component \citep{moritani11}. A gas stream from a dense part of the Be disk to the neutron star is implied.  \citet{yan12} found an anti-correlation between the optical brightness and the  $H\alpha$ intensity in 2009 observations. This indicated that a mass ejection event had taken place before the 2009 giant X--ray outburst. Near--IR monitoring during the 2011 February giant  outburst and X--ray quiescent phases, showed a  $\sim$12$\%$ reduction in the near--IR flux during the periastron passage of the neutron star \citep{naik11}. No changes in the spectra were found.

During the 1994 giant outburst, broad Quasi Periodic Oscillations (QPO) from 27 to 72 mHz were detected \citep{finger96} confirming the presence of an accretion disk.  In the 1989 March/April  giant outburst, two cyclotron  resonance scattering features were detected at 45 keV  and 100 keV \citep{kendziorra94}, from which a magnetic field of B$\sim$4$\times$10$^{12}$ G was inferred.  First cyclotron line studies of the recent outbursts of \f that took place in 2009/2010 have been presented by \citet{caballero11}. The cyclotron line energy evolution revealed no significant variation of the cyclotron line energy. High resolution grating spectroscopy of A\,0535+262 with \textit{Chandra} detected for the first time an absorption consistent with the presence of a highly ionized outflow \citep{reynolds10}, supporting the picture of a continued large scale outflow at large accretion rates but in a non-jet form. Gamma-ray observations of \f \,   with VERITAS at very high energies (VHE; E$\geqslant$100 GeV) were  analyzed by \citet{Acciari11} during the X--ray giant outburst in  2009 December. No VHE emission was evident at any time. They also examined data from the contemporaneous observations of \f\, from the Fermi/Large Area Telescope at high-energy photons (E $>$ 0.1 GeV) and failed to detect the source at GeV energies. 

In this work we present contemporaneous X--ray/optical observations of the BeXRB system A\,0535+26. We will interpret the data in terms of the Be-disk interaction with the neutron star companion.  We  include new optical observations from 2009 October  until 2011 March  (MJD 55082 to 55665), from the Spanish Astronomical Observatories of Sierra Nevada (OSN), University of Valencia (OAO) and  the Liverpool Telescope (LT; La Palma) as well as the Chinese Xinglong Station of National Astronomical Observatories (NAOC), and  simultaneous X--ray observations by the \textit{Fermi}/GBM.

\section{OBSERVATIONS and ANALYSIS}

In order to study the long-term X--ray  behavior  of  A\,0535+26 we have collected observations  from a variety of  sources.  X--ray flux measurements by distinct missions from 1970 until 1989 were obtained from \cite{giovannelligraziati92}. Earth occultation observations with \textit{CGRO}/BATSE were provided by the team\footnote{http://gammaray.msfc.nasa.gov/batse/occultation}  covering  the period from 1991 to 2000. \textit{RXTE}  data provided by the All Sky Monitor (ASM)\footnote{http://xte.mit.edu/ASM$\_$lc.html} team  and \textit{Swift}/BAT transient monitor results provided by the Swift/BAT team\footnote{http://heasarc.gsfc.nasa.gov/docs/swift/results/transients}, extend the monitoring up to date. 
X--ray timing measurements were obtained from \textit{RXTE} and GBM. Optical measurements were obtained from several sources, described below.
In addition, in this work we present results for the continuous  monitoring of \f \, by \textit{Fermi} GBM \citep{gbmpaper}  since 2008 June 11.   The  GBM is an all-sky instrument  sensitive to X--rays and gamma rays with energies between $\sim$8 keV and $\sim$40 MeV.  GBM includes 12 Sodium Iodide (NaI)
scintillation detectors and 2 Bismuth Germanate (BGO) scintillation
detectors. The NaI detectors cover the lower part of the energy range,
from 8  keV to about 1 MeV. The BGO detectors cover the energy range of
$\sim$150 keV to $\sim$40 MeV. Only data from the NaI detectors were used in the analysis presented here.

All the total-flux lightcurves for \f \, by  GBM were obtained using the Earth Occultation technique\footnote{http://heastro.phys.lsu.edu/gbm} on CSPEC data (128 energy channels every 4.096 s). Since GBM is not a pointed or imaging instrument, in order to determine fluxes for known sources, we measure the change in the count rate observed in the NaI detectors when the source enters or exits Earth occultation.   A detailed description of the Earth occultation technique is given in \citep{case11, wilson12}. Here we give  a brief summary. For each occultation step,  the count rates in  a 4--minute window centered on the occultation time are fitted with a quadratic background plus source terms for the source of interest and each interfering source that appears in the window. The source terms consist of an energy dependent atmospheric transmission model and a time-dependent model count rate, derived from the time-dependent  detector response convolved with an assumed source spectrum.   Each energy channel and each detector is fitted independently. For each source term a scaling factor is fitted, along with the quadratic background coefficients. When multiple detectors are included in the fit, the weighted mean for the scaling factor of the source of interest is computed for each energy channel. The mean scaling factor is then multiplied by the predicted flux in each energy channel to obtain flux measurements for the source of interest.  In the case of A\,0535+26, the assumed spectrum \citep{kendziorra94} consists of a power law with a high energy cutoff and  a Lorenzian cyclotron absorption model.  \citep{case11} and \citep{wilson12} find that flux values do not depend strongly on the choice of assumed model. 

\begin{longtable}{lllll}
  \caption[Recent H$_\alpha$ EW and V magnitudes]{Recent H$_\alpha$ EW  (error $\pm$0.5) and V magnitudes.} \\\hline
  \label{v_alpha} \centering
  MJD        &  H$_\alpha$  EW            &    V mag   & err   &  Telescope \\\hline
  \endfirsthead
\caption{continued from previous page}\\\hline  
     MJD        &  H$_\alpha$  EW            &    V mag   & err   &  Telescope \\\hline
\endhead
55060.50  &	    &  8.900  &  0.05	  & OAO\\ 
55082.00 &  -20.7  &           &	&   LT\\
55083.00 &  -21.3  &           &	&   LT\\
55084.50 &  -21.6  &           &	&   LT\\
55087.00 &  -21.5  &           &	&   LT\\
55087.50 &  -21.0  &           &	&   LT\\
55091.00 &  -21.0  &           &	&   LT\\
55115.50  &	    &	9.120  &  0.01 &  OSN\\ 
55116.50  &	    &	9.110  &  0.01 &  OSN\\ 
55117.50  &	    &	9.100  &  0.01 &  OSN\\ 
55127.88  &  -24.70 &         &        &  NAOC  \\
55128.50  &	    &  9.150  &  0.01	&  OSN\\ 
55128.84  &  -24.82  &       &        & NAOC\\
55129.50  &	    &  9.140  &  0.01	&  OSN\\ 
55129.83 &  -24.38  &         &        & NAOC  \\
55130.82 &  -25.58  &        &        &  NAOC \\
55131.50 &	    &  9.150     &  0.01  & OSN\\ 
55132.85 &  -25.02  &	  &	   &   NAOC  \\
55133.82 &  -25.86  &	  &	   &   NAOC  \\
55145.00 &  -23.6  & 	    &	     &   LT  \\
55147.00 &  -23.5  & 	    &	     &   LT  \\
55150.00 &  -23.6  & 	    &	     &   LT  \\
55154.00 &  -24.6  & 	    &	     &   LT  \\
55158.00 &  -24.7  & 	    &	     &   LT  \\
55162.00 &  -24.0  & 	    &	     &   LT  \\
55166.00 &  -24.5  & 	    &	     &   LT  \\
55168.00 &  -23.9  & 	    &	     &   LT  \\
55171.00 &  -23.4  & 	    &	     &   LT  \\
55175.00 &  -23.3  & 	    &	     &   LT  \\
55176.81 &  -23.14 & 	    &	     &  NAOC \\
55191.00 &  -23.6  & 	    &	     &   LT  \\
55196.00 &  -23.7  & 	    &	     &   LT  \\
55199.00 &  -24.2  & 	  &	   &	LT \\
55203.00 &  -23.8  & 	  &	   &	LT \\
55208.00 &  -23.5  & 	  &	   &	LT \\
55211.00 &  -23.2  & 	  &	   &	LT \\
55213.00 &  -21.9  & 	  &	   &	LT \\
55214.50 &	  &   9.230   &  0.01  & OSN\\ 
55218.00 &  -22.2 & 	 &	  & LT   \\
55220.00 &  -22.6 & 	 &	  & LT   \\
55223.90  &  -23.0  &     &      & LT  \\
55309.00  &  -22.0  &     &      & LT  \\
55309.90  &  -22.0  &     &      & LT  \\
55487.00  &  -11.0  &     &      & OSN  \\
55494.00  &  -11.8  &     &      & LT  \\
55495.10  &  -11.8  &     &      & LT  \\
55496.10  &  -12.1  &     &      & LT  \\
55499.20  &  -11.9  &     &      & LT  \\
55502.10  &  -10.4  &     &      & LT  \\
55505.10  &  -10.3  &     &      & LT  \\
55508.20  &  -10.7  &     &      & LT  \\
55513.10  &  -11.5  &     &      & LT  \\
55516.00  &  -12.1  &     &      & LT  \\
55517.10  &  -12.3  &     &      & LT  \\
55520.20  &  -12.4  &     &      & LT  \\
55523.10  &  -11.1  &     &      & LT  \\
55542.42  &	    &   9.269    &  0.02  & OSN\\ 
55544.00  &  -10.5  &     &      &	 LT  \\
55545.10  &  -10.5  &     &      &	 LT  \\
55545.52  &	   &    9.285  &  0.02  & OSN\\ 
55546.44  &	   &    9.293  &  0.02  & OSN\\ 
55554.20  &  -11.3  &     &        & LT  \\
55564.00  &  -10.2  &     &        & LT  \\
55565.10  &  -10.2  &     &        & LT  \\
55572.90  &  -9.3  &      &        & LT  \\
55574.00  &  -9.2  &      &        & LT  \\
55574.00  &  -9.5  &      &        & OSN  \\
55574.90  &  -9.2  &      &        & LT  \\
55578.90  &  -10.1  &     &        & LT  \\
55580.00  &  -10.0  &     &        & OSN  \\
55580.10  &  -10.5  &     &        & LT  \\
55595.52 &	  &    9.271  &  0.02  & OSN\\ 
55599.46 &	  &    9.272  &  0.02  & OSN\\ 
55600.48 &	  &    9.266  &  0.02  & OSN\\ 
55601.45 &	  &    9.269  &  0.02  & OSN\\ 
55602.44 &	  &    9.276  &  0.02  & OSN\\ 
55603.51 &	  &    9.268  &  0.02  & OSN\\ 
55604.44 &	  &    9.263  &  0.02  & OSN\\ 
55605.43 &	  &    9.265  &  0.02  & OSN\\ 
55608.90  &  -10.0  &     &      & LT  \\
55610.90  &  -11.0  &     &      & LT  \\
55614.90  &  -12.0  &     &      & LT  \\
55622.90  &  -10.8  &     &      & LT  \\
55664.90  &  -9.2  &      &      &LT  \\\hline
\end{longtable}

\begin{figure*}[!t]
\hspace{-0.45cm}
 \includegraphics[width=17.25cm,height=15cm]{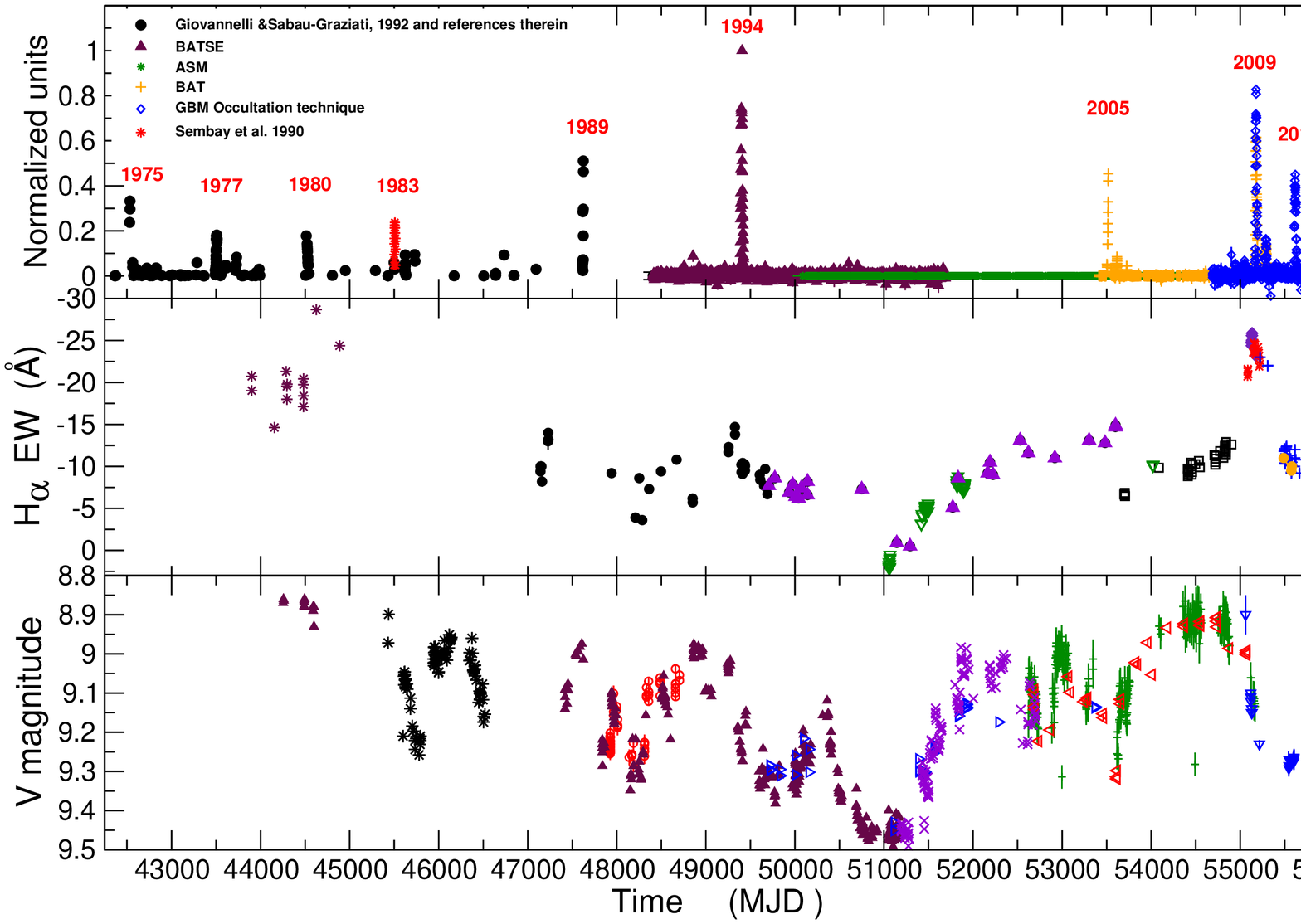}
\caption{Top. History of the X--ray outbursts undergone by A\,0535+26 in arbitrary units. The peak-intensities of the different outbursts are only  intended to be illustrative  of the times and types of events, since they have not been corrected for the different energy bands.  Middle.\, Long-term  history of the evolution of the H$_\alpha$ equivalent width. Bottom. Evolution of the visual magnitude.  The compilation of H$_\alpha$ EW and visual magnitudes come from different authors described in the text (see Sect~\ref{color}).}
\label{long_term_flux}
\end{figure*}

Large deviations from expected fluxes were only seen if the model is very inconsistent, e.g. a soft model with a rapid turn-over for a hard source or a very hard model for a soft source. Since each energy channel is fitted independently and channel to channel variations are not constrained by the model, hardness ratios also appear not to strongly depend upon the choice of spectral model. A\,0535+26 was not detected in the NaI detectors above 100\,keV, even in the 10 days around the giant outburst peak. Since the NaI detector area is comparable to or larger than the BGOs (depending upon the combination of NaI detectors observing the source), the BGOs were not used.

Timing analysis was carried out with GBM  CTIME data (with  8 channel spectra every 0.256 seconds) from channels 0, 1, 2 and 3 ($\sim$8-12\,keV, $\sim$12-25 keV, $\sim$25-50 keV  and $\sim$50--100 keV, respectively).   After the appropriate NaI detector rates (from all 12 detectors and the 4 lowest channels) are selected,  an empirical background model is fit and subtracted.  The residuals to this, which contain the pulsations of the pulsars, are used to obtain  pulse frequencies, pulse profiles and pulsed fluxes. Then, a pulsed search was made over a narrow frequency band. A detailed explanation of our technique can be found in \cite{finger09} and \cite{camero10}.

The Rossi X--ray Timing Explorer ({\it RXTE}) \citep{bradt93} carries 3 instruments on board. The Proportional Counter Array (PCA) \citep{Jahoda96} is sensitive from 2--60 keV. The High Energy X--ray Timing Experiment (HEXTE) \citep{Gruber96} extends the X--ray sensitivity up to 200 keV.  Monitoring the long-term behavior of some of the brightest X-ray sources, the All Sky Monitor \citep{levine96} scans most of the sky every 1.5 h at 2-10 keV\textit{RXTE}  observed \f\,  since 2005. Recent observations since the beginning of 2009 have been analyzed in this work. For the pulse profile study  and for each available observation, we have analyzed either GoodXenon  or event mode PCA data (time bin size 125µs, 64M energy channels) using Ftools V6.9. The available  PCA pulse profiles obtained in the $\sim$8--14\,keV and $\sim$14--21\,keV energy bands are consistent with the GBM pulse profiles from channels 0 and 1.

Regarding the optical data,  we include photometric archival observations from HIPPARCOS, ASAS, \textit{INTEGRAL}/OMC, and recent ground based observations from the 0.9 and 1.5\,m telescopes at the OSN, and the 60\,cm telescope at the OAO.  We transformed the HIPPARCOS filter to Visual magnitude,  following the procedure found in \cite{harmanec98} and applying B--V and U--B values from \cite{Lyuty00}. We computed the V magnitude from \cite{coe06}  based on the work done by \cite{grundstrom07}.  As we mentioned before, the spectroscopic observations come from the 2\,m telescope at LT  and the 2.16\,m telescope at NAOC. Standard tools and photometry packages in IRAF were used for the reduction and analysis of the new data set of  observations.

\begin{figure*}[!]
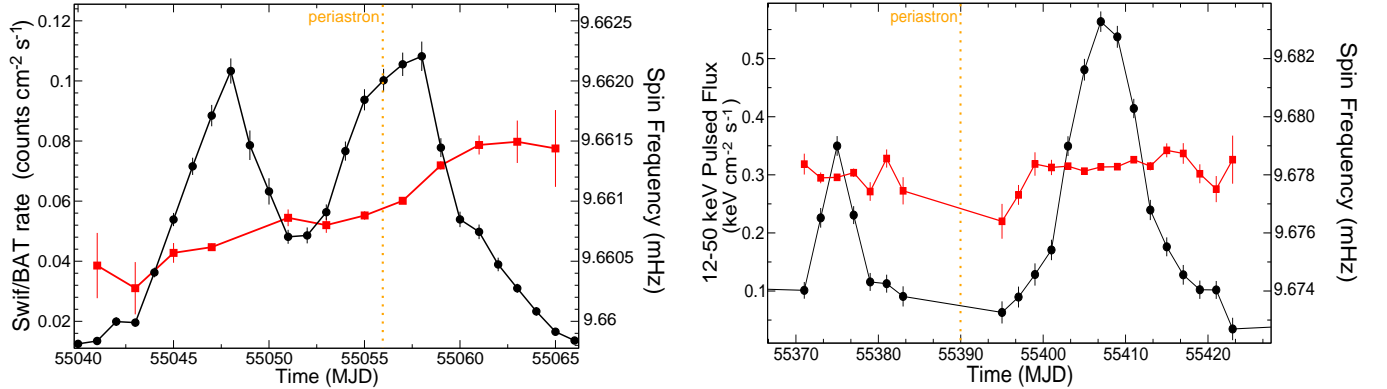

 \includegraphics[width=8.75cm,height=5.15cm]{figure2_1.eps}
 \hspace{0.25cm}
  \includegraphics[width=8.75cm,height=5.15cm]{figure2_2.eps}
\caption{Right. BAT lightcurve of the double-peaked normal outburst  occurred in  2009  August (black circles).  Overplotted is the spin frequency history during the event by  GBM (red squares). Right.  GBM 12-25 keV pulsed flux vs. time for the 2010 July normal outburst, with the frequency history also overplotted.}
\label{2xnormal}
\end{figure*}

\section{OUTBURSTS HISTORY}

\subsection{Long-term History}

Figure~\ref{long_term_flux}  shows a long-term X--ray/optical overview of A\,0535+26/HDE\,245770. The pre-BATSE  X--ray flux measurements found in \cite{giovannelligraziati92} come from a large variety of instruments, operating in different energy bands, and were presented  only in Crab units.  Due to the fact that the conversion of those measurements to real flux values is not straightforward and in order to show a global picture of this system, all the X--ray flux/count-rates were converted in to Crab units and then normalized using the maximum value for the whole data set (that is, BATSE). Therefore the peak-intensities of the different outbursts are intended to be illustrative  of the times and types of outbursts. This will allow direct comparisons with the optical data and help us to understand the underlying Be-neutron star interaction.  

The typical outburst peak luminosity for \f  during a  normal outburst  is  $\sim$400 mCrab and between 3--6 Crab  for a giant one, in the 15--50\,keV energy range. In the upper panel of  Figure~\ref{long_term_flux} we see that after the giant outbursts in 1975 (the discovery), 1977, 1980, 1983 \citep{sembay90} and 1989,  BATSE detected this source in 1994 \citep{bildsten97}.   The 1994 giant outburst was preceded by three normal outbursts and followed by two small normal outbursts. Then  A\,0535+26 went back to quiescence for almost 11 years.  Only  a few observations were reported during that period \citep{coe06, hill07,negueruela00,orlandini04}.   The source renewed activity in 2005 May with another giant outburst and  two normal outbursts. Each of those normal outbursts began approximately a week before the periastron passage in the A\,0535+26$'$s 111 day orbit. Unfortunately the 2005 giant outburst was poorly observed due to the closeness of the Sun to the source. There are only a handful of \rxte\,/ASM dwells and serendipitous {\it Swift}/BAT observations near the outburst peak. In addition a pointed RHESSI observation \citep{smith05} has been reported.

\subsection{Recent Outbursts}

After almost three years of being in quiescence \f\, was again active in 2008 undergoing  four consecutive normal outbursts plus a giant outburst in 2009 December (see Figure~\ref{long_term_flux}).  The very large event occurred in  2009 December was foreseen due to the fact that new X--ray activity was detected 19 days before periastron.  Unexpectedly, another outburst peaking at $\sim$1 Crab (in the 15-50\,keV band) occurred, with intermediate low X--ray activity being detected. This event was followed by 2 weak normal outbursts.  In 2011 February  another giant outburst took place, approximately 11 days before the next periastron passage (2011 February 20). The outbursts of higher fluxes began at an earlier orbital phase.  It is to be noted that this is the shortest interval in which two (or maybe three?) giant outbursts have been reported from \f. \\

\subsubsection{Some Peculiar Normal Outbursts}

Initial flaring of an outburst in 2005 September was seen as a double peak by \textit{RXTE}.   The highest flux in this outburst  was reached during the short spike prior to periastron passage, and it was followed by  a more gradual rise to a peak and then an exponential tail  \citep{finger06}.  Similar pre--periastron peaks have been seen in other Be/X--ray binaries such as 2S\,1845-024 and EXO\,2030+375 \citep{camero05, wilson08}.   Using \textit{Swift}/BAT data \citet{postnov08} found that this was a regular normal outburst but exhibiting large flaring activity. Particularly intense were a series of small flares taking place at the beginning of this event  compared to the rest of the outburst.  We also want to note that the remaining part of the outburst, as well as other outbursts of the source, present a series of small flares instead of a smooth evolution of the flux with time. 

More recently, \f\,  underwent two  double peaked normal outbursts in 2009 August and 2010 July (see Figure~\ref{2xnormal}). We point out that the one in 2009 August  showed  two approximately equal peaks in both duration and intensity  (see left panel of Figure~\ref{2xnormal})  without the typical shape of pre-periastron peaks described before. The last peculiar outburst in  2010 July  was only entirely measured by GBM. One could note that this outburst is also noticeable in the BAT data, but these unfortunately  have a gap up to the first peak (MJD 55374) because of the Sun (on 2010 June 25, the Sun was $\sim$8.5$^\circ$ from A\,0535+26).  Once more we found no resemblance with  previous double-peaked outbursts. The first peak was smaller, shorter  and totally detached from the main outburst peak (see right panel of Figure~\ref{2xnormal}).  During  the 2005 pre-outburst phase  the pulse profile changed dramatically \citep{camero07,caballero07}, as well as the value of the center of the cyclotron line energy \citep{caballero08}. However in the 2009 and 2010  normal outbursts no changes have been found neither with \textit{RXTE}/PCA nor GBM (see Sect~\ref{disccus_prof}).  \\

\begin{figure}[]
\hspace{-0.5cm}
 \includegraphics[width=9.5cm,height=7.5cm]{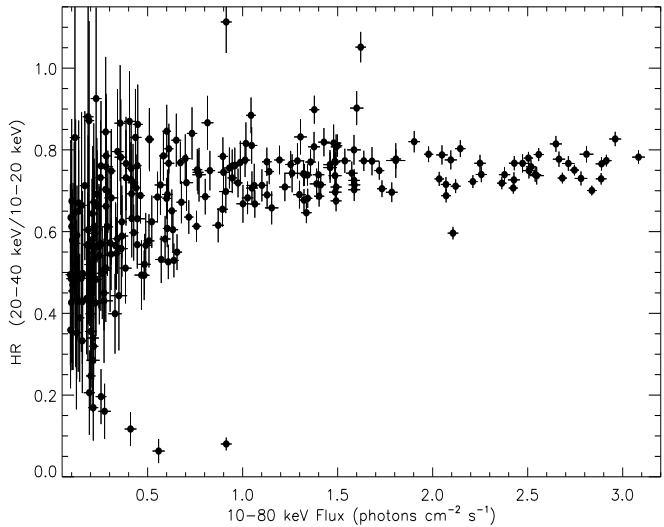}
 \caption{Hardness ratio analysis using GBM Earth Occultation data  for \f\,. The vertical axes indicates the flux  in the 10--80\,keV band. The HR has been defined by 20--40keV\,keV/10--20\,keV. The presence of observations with lower hardness ratio seems to point out  a
change in the spectral continuum, i.e.,  a spectral cut-off at higher energies or an increased contribution from the softer components related to the neutron star and boundary layer.}
\label{hr}
\end{figure}


\begin{figure*}[!t]
\hspace{-1.5cm}
\includegraphics[width=19.25cm,height=20cm]{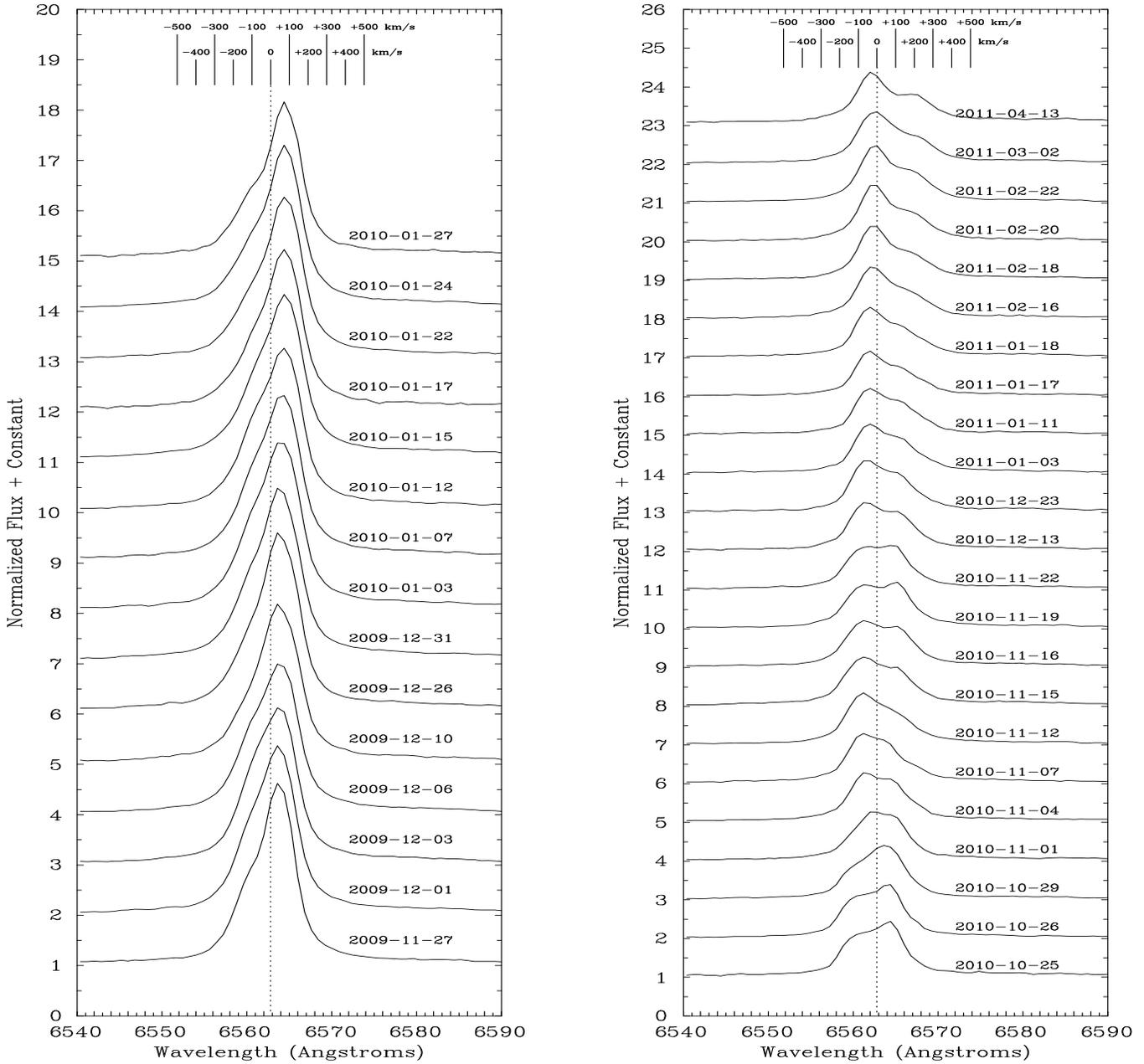}
\vspace{-0.3cm}
 \caption{Recent H$_{\alpha}$ line profile evolution from 2009 October up to 2011 March. All spectra have been normalized by dividing by the mean flux calculated between 6450\,{\AA} and 6500\,{\AA}. As a reference, some of the most recent outburst peaked  around  2009 August 7 -- December 17(giant), 2010 March 28 -- July 1 -- September 1  and 2011 February 28 (giant).} \label{halfa_prof}
\end{figure*}

 
\begin{figure*}[!t]
\vspace*{0.15cm}
 \includegraphics[width=18.15cm,height=13cm]{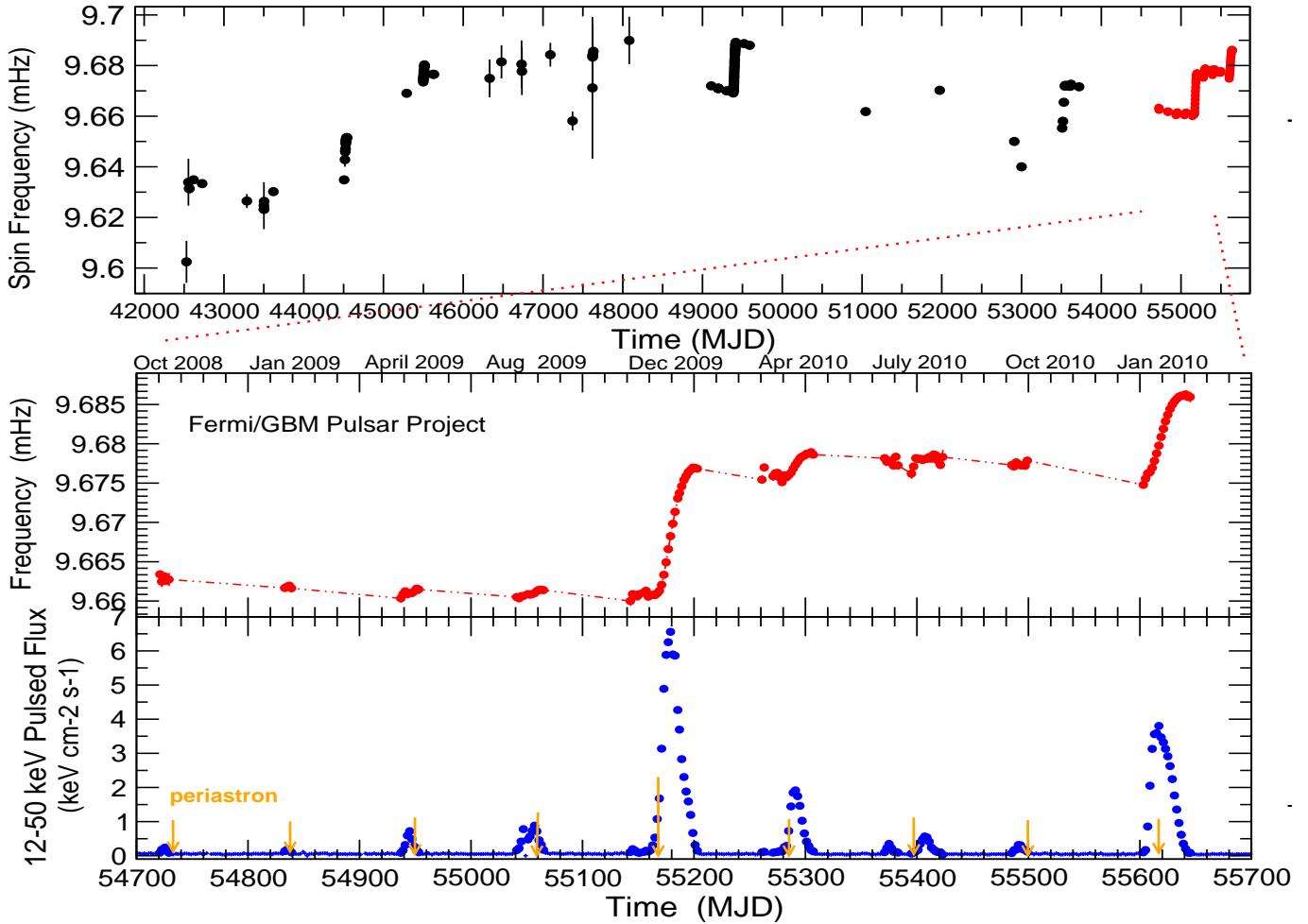}\vspace*{0.1cm}
 \caption{Top. Long-term frequency history of A\,0535+26 since 1975 (see text for references). Middle. Zoom of the frequency history of this source since 2008 by GBM. The frequencies are connected with a dot-dash line. Bottom: Daily average 12--50 keV pulsed flux measured with GBM  }
\label{long_term_freq}
\end{figure*}

\subsubsection{Spectral  Hardness}

Fig~\ref{hr} shows the hardness ratio--intensity (HRI) diagram analysis carried out for this source. This allows us to study  the   spectral variability of  \f\,   without the supposition of any spectral model.   GBM total flux lightcurves  from 10 keV up to 80 keV (5\,keV and 10\,keV wide) were obtained using the Earth Occultation technique.  The HR was defined  as  20--40\,keV/10--20\,keV and we used the flux in the 10--80 keV band for the intensity in our diagram. To reduce large uncertainties the light curves were rebinned and then the HR were computed. 
It is to be noted that like all hardness ratios, these are instrumentally dependent.  From  that figure  we can see that the vast majority of observations reveal a constant hardness ratio. Only at low flux intensity levels there seems to be a smooth hardening of the spectrum of \f\,as the flux intensity increases.  Therefore the diagonal and horizontal  branches in the HRI diagram observed by \citet{reig08} in a correlated spectral-timing analysis of four Be/X-ray binaries, are not present for A\,0535+26.  We point out that the study by \citet{reig08}   was carried out during major X-ray outbursts and not all sources displayed  the entire pattern of variability.

 \subsection{Optical Measurements}\label{color}
 
The middle and bottom panels of Figure~\ref{long_term_flux} show the H$_\alpha$ EW and V magnitude long-term evolution for HDE\,245770 since $\sim$1978. The compilation of H$_\alpha$ EW measurements come from \citet{aab85}(stars), \citet{clark98} (filled circles), \citet{coe06} (filled up triangles), \citet{grundstrom07} (down triangles) and \citet{moritani10} (squares). The historical observations of the V magnitude come from \citet{gnedin88} (stars), Lyuty and Zaitseva (2000) (up triangles), \citet{zaitseva05} (crosses) and \citet{coe06} (right triangles), as well archived observations from HIPPARCOS (circles), ASAS3 (plus), \textit{INTEGRAL}/OMC (left triangles). Table~\ref{v_alpha} shows recent  measurements of the  H$_\alpha$ EW and the V magnitudes obtained with  LT , NAOC, OSN and  OAO.
 

In Figure~\ref{long_term_flux}  we see that, in the long-term, both the H$_\alpha$ EW and the V magnitude  show a parallel global trend.
From MJD $\sim$44000 to 50000 both magnitudes presented an overall decreasing trend, and after the minimum is  reached they climbed back up to  2009 November ($\sim$\,MJD 55136) .  Then,  both magnitudes peaked in this period, and  afterwards  a dramatic decrease took place which continues through 2011 March.   Furthermore, the shape of the H$_{\alpha}$ profile was more  simple around 2009, evolving from a clear double structure \citep{coe06} to a single peak.  Later on,  a  double peaked structure is observed (see Fig.~\ref{halfa_prof}).

\section{TIMING ANALYSIS}

We have performed distinct  types of timing analysis for \f\,  in X--rays. These include a  study of the evolution of the pulse frequency, a search for Quasi Periodic Oscillations and a  pulse profile study.

\subsection{Long-term Pulse Frequency Evolution}

The top panel of  Figure~\ref{long_term_freq}   shows a long-term pulse frequency history for A\,0535+26 obtained by several authors since 1975 to 2006. Among them:  \citet{rosenberg75}, \citet{fishman77}, \citet{bradt76}, \citet{li79},  \citet{hameury83}, \citet{frontera85}, \citet{nagase82}, \citet{sembay90}, \citet{motch91}, \citet{coe90}, \citet{makino89}, \citet{sunyaev89}, \citet{cusumano92}, \citet{finger94}, \citet{finger06} and \citet{finger09}.

\f\,  shows a  brief spin-down period (MJD 42000--44000) followed by a global spin-up trend (MJD 44000--49500), and  ending  with a spin-down trend. Although between MJD 50000 to 53500,  no X--ray outburst activity was reported from this source (see Fig. ~\ref{long_term_flux}), \citet{coe06} detected a 110$\pm$0.5 d orbital modulation pattern during this long  quiescent X–-ray state with \textit{RXTE}/ASM.  
\citet{hill07},  \citet{negueruela00} and \citet{orlandini04}  detected  pulsations at $\sim$103.5\,s  with different instruments during this period. Then, suddenly the 2005 giant outburst took place and \f\, exhibited a rapid spin-up.  The following giant outbursts in 2009, 2010 and 2011 allowed \f\,  to slowly recover from the overall spin-down period. We note that the transitions between giant outbursts resulted in  global smooth spin-down periods.  Nevertheless, in spite of  this overall picture  \f\, might be just exhibiting a random walk in pulse period over the long-term.

\subsection{Pulse Frequency Evolution Monitored by \textit{Fermi/GBM}}

The middle panel of Figure~\ref{long_term_freq} shows the recent pulse frequency history by GBM.
We have used the orbital elements from \cite{finger96} with adjusted epoch and period, i.e.
 P$_{orb}$=111.07(7) days and  T$_{periastron}$= MJD 53613.00(16). After the first normal outburst detected by GBM in 2008 October,  \f \, experienced an overall steady spin-down trend  up to the 2009 December  outburst. Then, a  rapid spin-up  occurred followed by a return to a steady spin-down trend. This seems to be a common pattern for A\,0535+26 also seen surrounding another giant outbursts. The bottom panel of the same Figure shows the 12--25 keV pulsed flux for the same period  by GBM. Once again, we find a strong pulsed flux/spin-up rate correlation, indicating the presence  of an accretion disk around the neutron star.

\subsection{Quasi--Periodic Oscillations in the 2009 December Giant Outburst}

\begin{figure}[!t]
\hspace{-0.5cm}
\includegraphics[width=9cm,height=10.5cm]{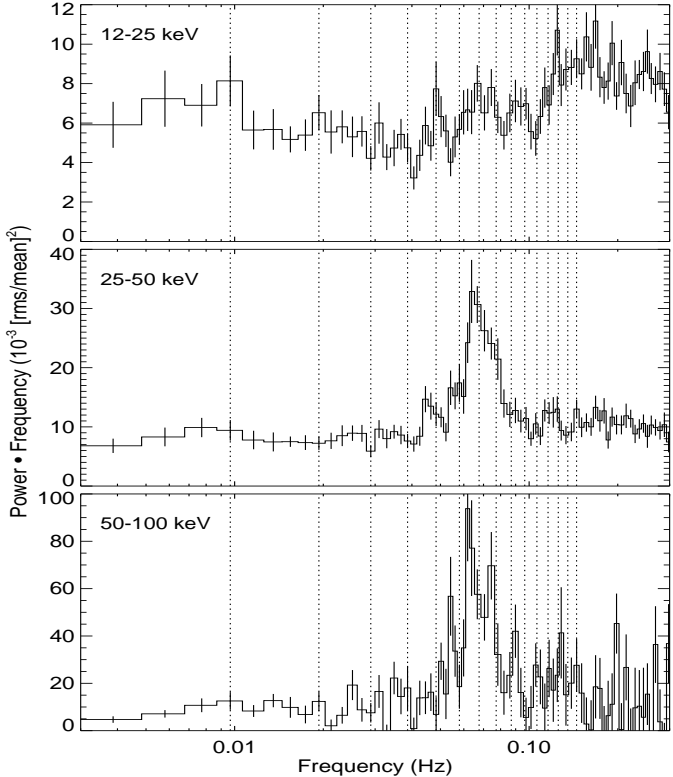}
\caption{QPO centered at 62 mHz from a GBM observation on 2009 December 11  in three energy bands. We can clearly see that this QPO is stronger in the 50--100 keV band but not detected in  12--25 keV range. Dotted lines denote pulse harmonics. The daily mean pulse profile was subtracted from the data before power spectra was made.}\label{qpo}

\end{figure}

\begin{figure*}[!t]
\hspace{2.7cm}
\includegraphics[width=10cm,height=8cm]{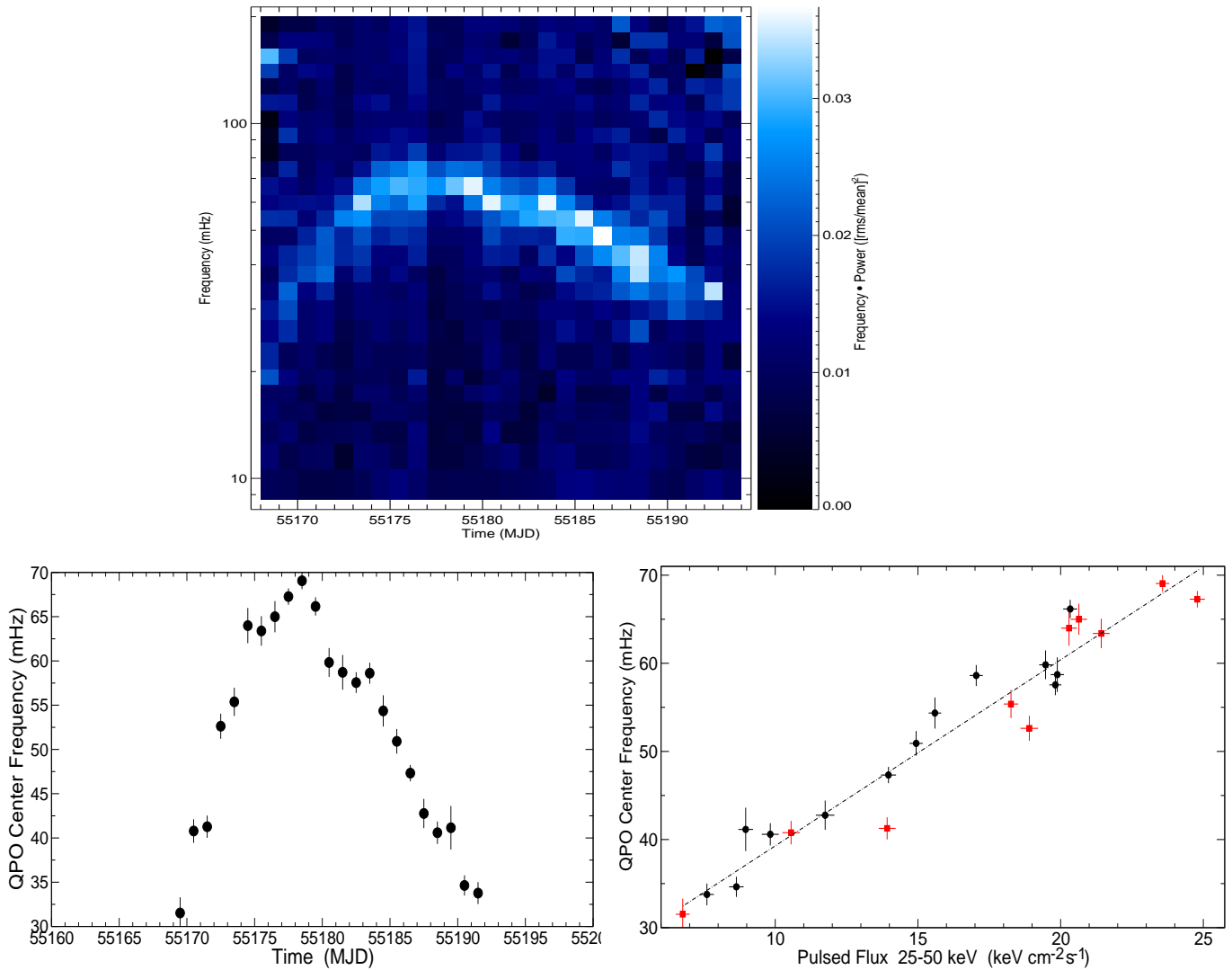}\\\\
\includegraphics[width=9cm,height=6cm]{figure7_1.eps}
 \hspace{-0.15cm}
  \includegraphics[width=9cm,height=6cm]{figure7_2.eps}
\caption{Top. Dynamic power spectra for A\,0535+26 showing the evolution of the QPO center frequency in the 25--50keV band.  The color scale (see online version for colors) gives the fractional power per logarithmic frequency interval. Bottom. QPO center frequency evolution vs. time (left) and vs. the  pulsed flux in the 25-50\,keV band (right). A  strong QPO center frequency/X--ray flux correlation is found. Red points indicate the rise of the outburst and  black the fall. The best linear fit gives a correlation coefficient of 0.9677762 (dashed line).\label{qpo_evol}}
\end{figure*}

During the 1994 giant outburst, a broad QPO from 27 to 72 mHz was detected with BATSE \citep{finger96} in the 25-60\,keV band.  With GBM we analyzed  the aperiodic variability in the X--ray flux of A\,0535+26 following the method described in \cite{finger96}. During the 2009 December giant outburst we found that the power spectra of the hard--X ray flux of this source,  between 3 mHz and 1 Hz, consisted of an approximately 1/f power law continuum plus a broad QPO  and a pulse component \citep{fingerwilsoncamero2009}.  On December 10 the QPO was centered on 62$\pm$1 mHz (FWHM of 29$\pm$2 mHz) in the 25--50 keV band. Figure ~\ref{colorscale}  shows  a  color-scale representation of the evolution of the power spectra over the course of the outburst in the 25--50 keV band. The QPO is evident during much of the outburst, and shows a marked evolution in centroid frequency. We were able to detect the QPO from December 4  to 27.   In Fig.~\ref{qpo} we can see that this QPO was stronger in the 50--100 keV band, but not detected in the 12--25 keV range\footnote{We note that this is the most sensitive energy band in GBM.  Further studies investigating these results will be carried out in a forthcoming paper using data from other X-ray missions, as for instance \textit{RXTE} and \textit{Suzaku}.}.  In the bottom-left panel of Fig.~\ref{qpo_evol}  it is also shown how the center frequency rose from 30 to 70 mHz and went back to 30 mHz. In addition a  strong QPO center frequency/X--ray flux correlation is found (bottom-right panel of Fig.~\ref{qpo_evol}). When we perform a linear fit to the QPO center frequency and X--ray flux data sets we obtained for the best fit a correlation coefficient of 0.9677762.

\begin{figure*}[!]
\includegraphics[width=6cm,height=12.0cm]{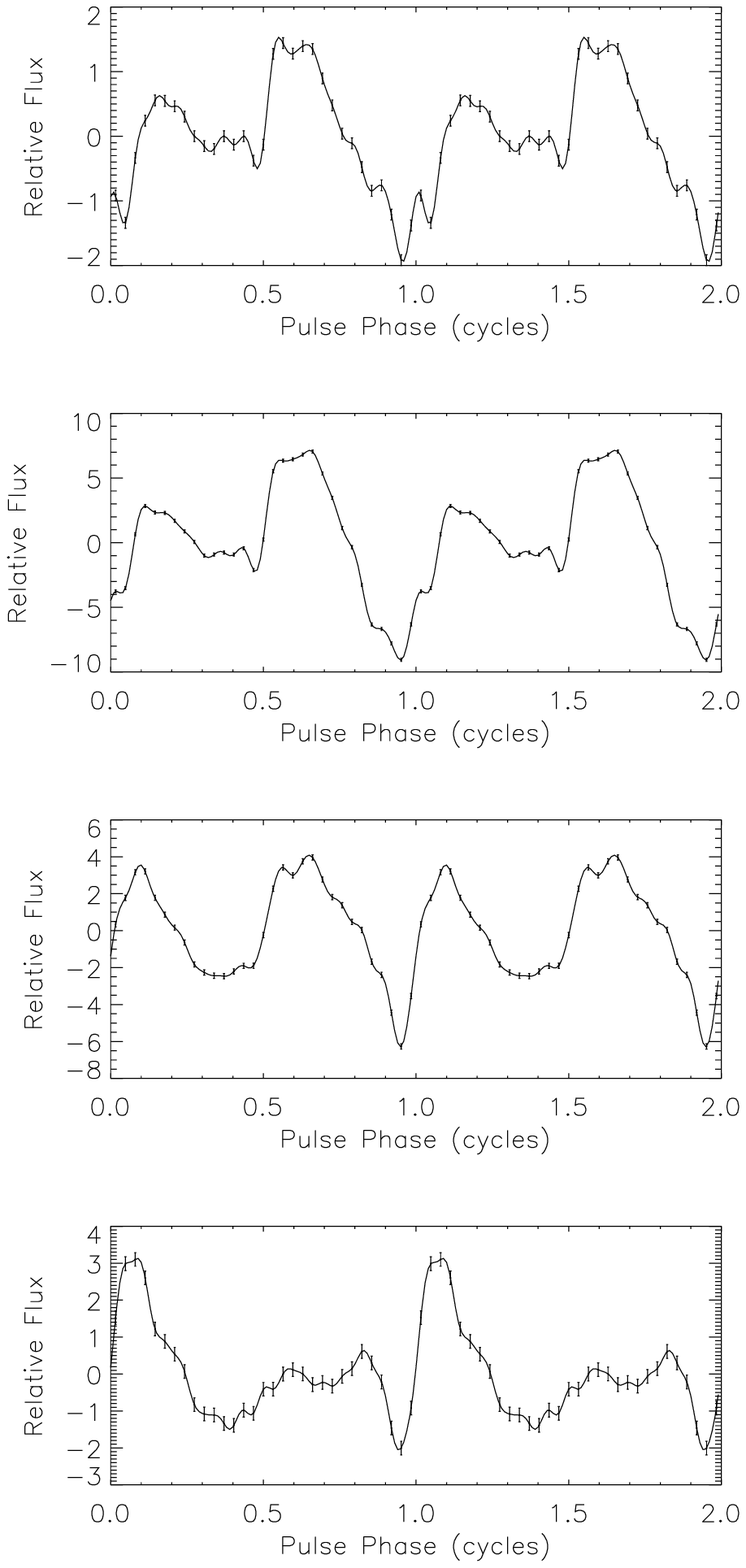}
\includegraphics[width=5cm,height=11.9cm]{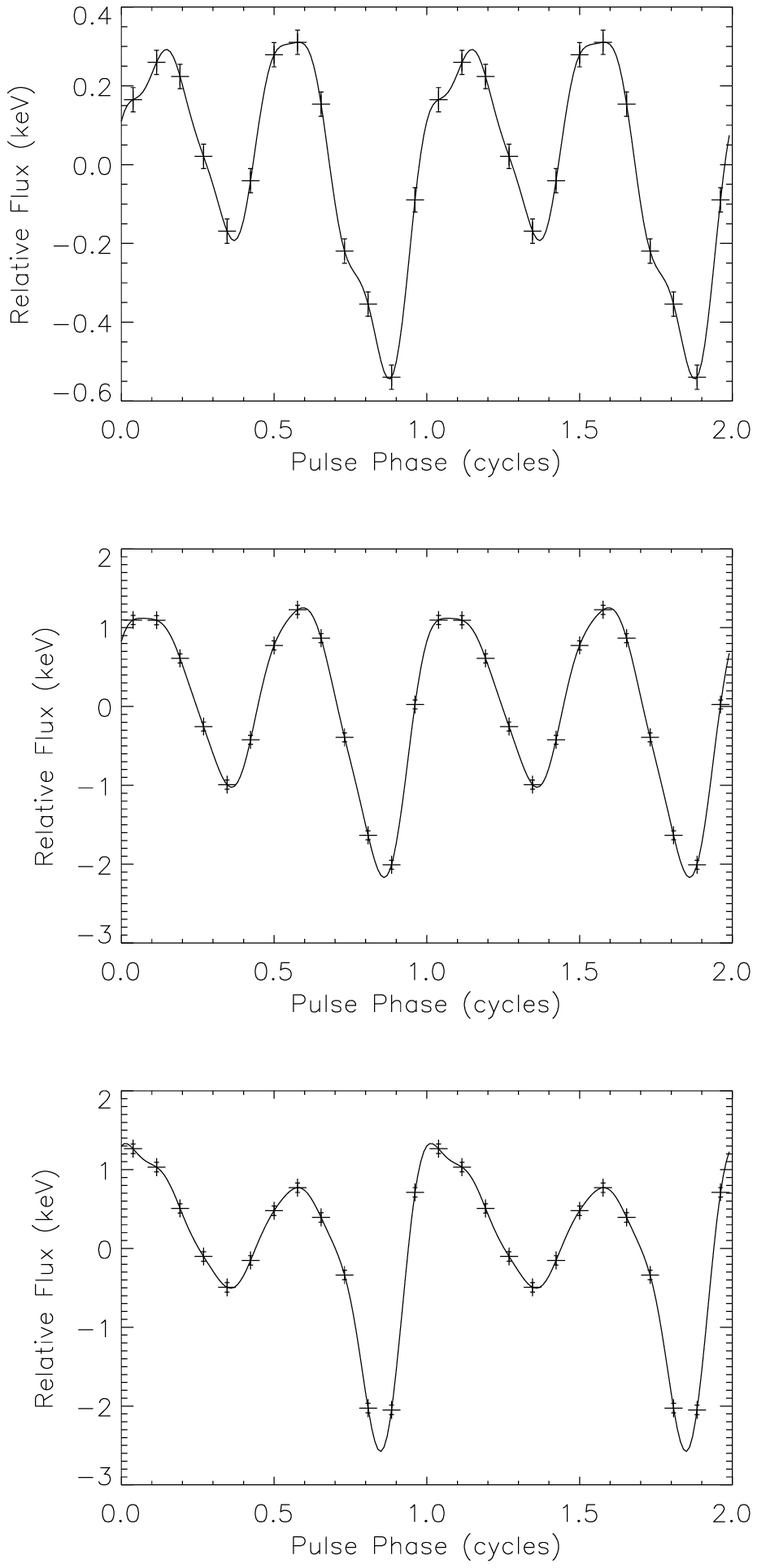}
\includegraphics[width=5cm,height=11.8cm]{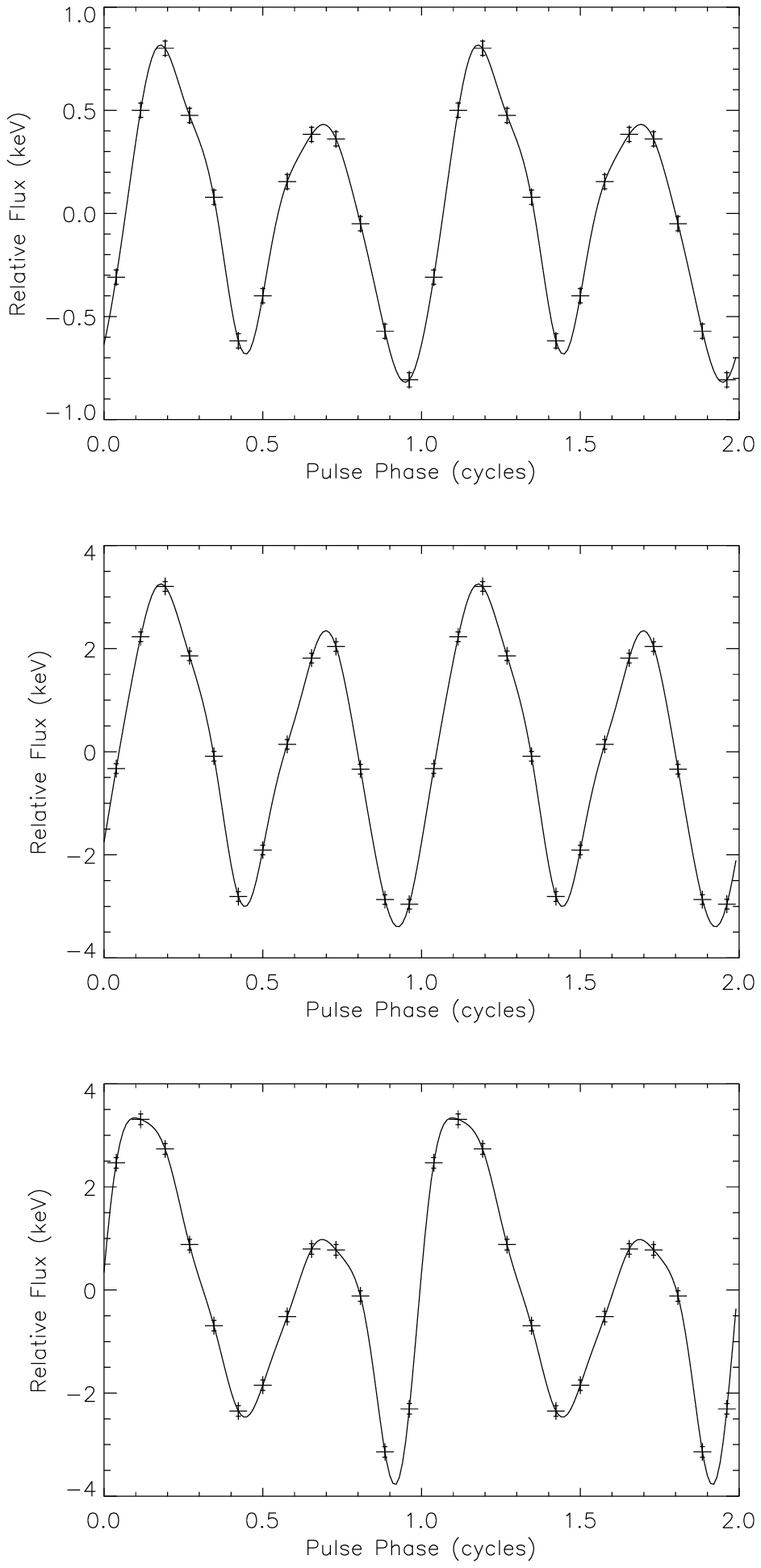}
\caption{Left. Pulse profile correlation with energy  during the 2009 December giant outburst near the maximum (from top to bottom: 8--12, \,20--25,\,25--50, and 50--100\,keV bands) by GBM. Two cycles are shown for clarity. Middle. Pulse profiles near the maximum of the large event in 2010 March (from top to bottom: 8--12, \,20--25,\,25--50\,keV bands). Right. Pulse profiles of the giant outburst in 2011 March in the same bands.}\label{prof_giant}
\end{figure*}


\begin{figure*}[!t]
\includegraphics[width=15cm,height=8cm]{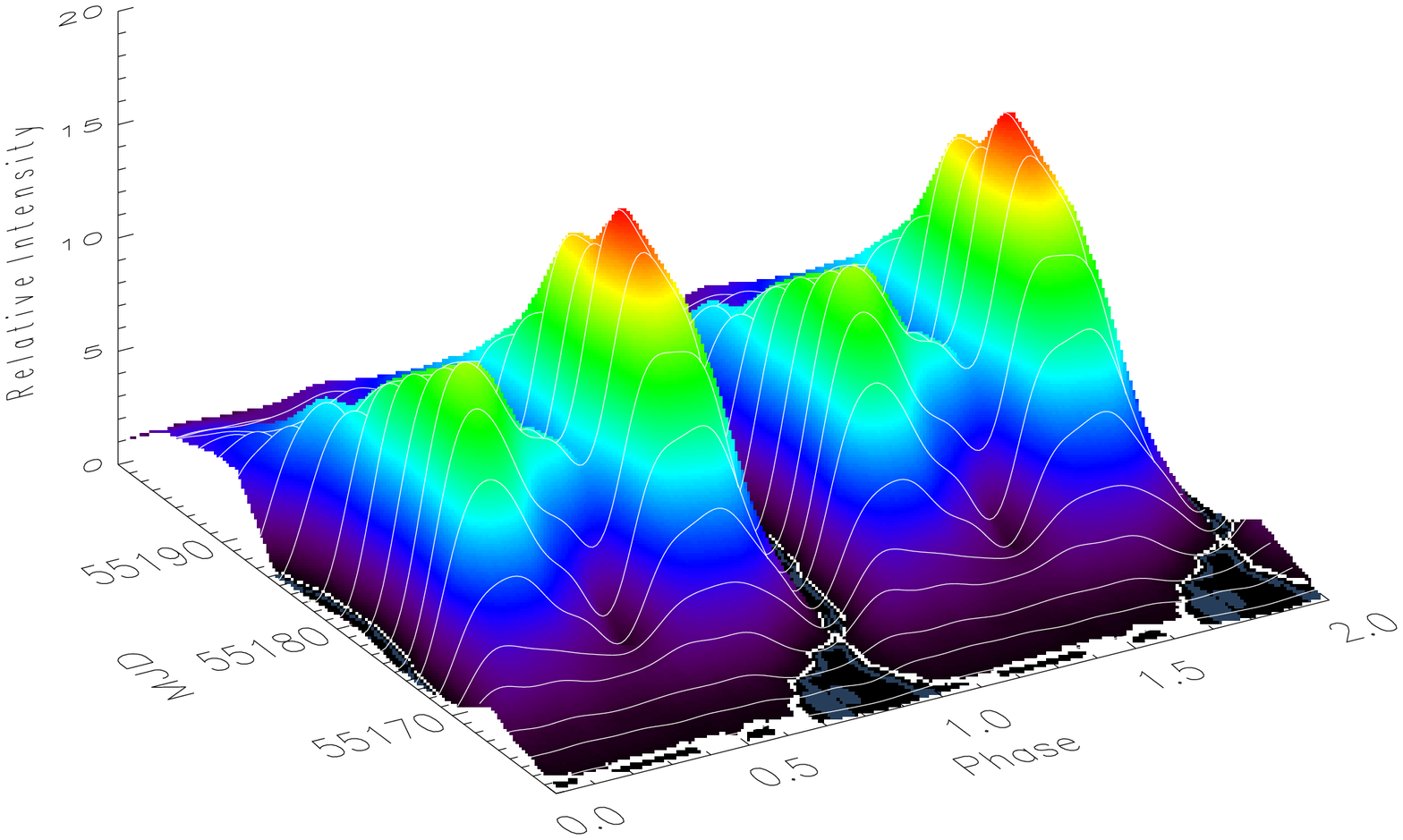}

\includegraphics[width=1cm,height=6cm]{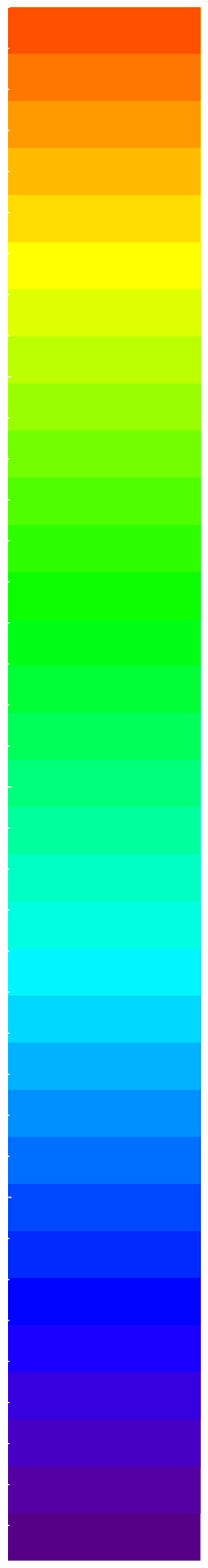}
\includegraphics[width=14cm,height=7cm]{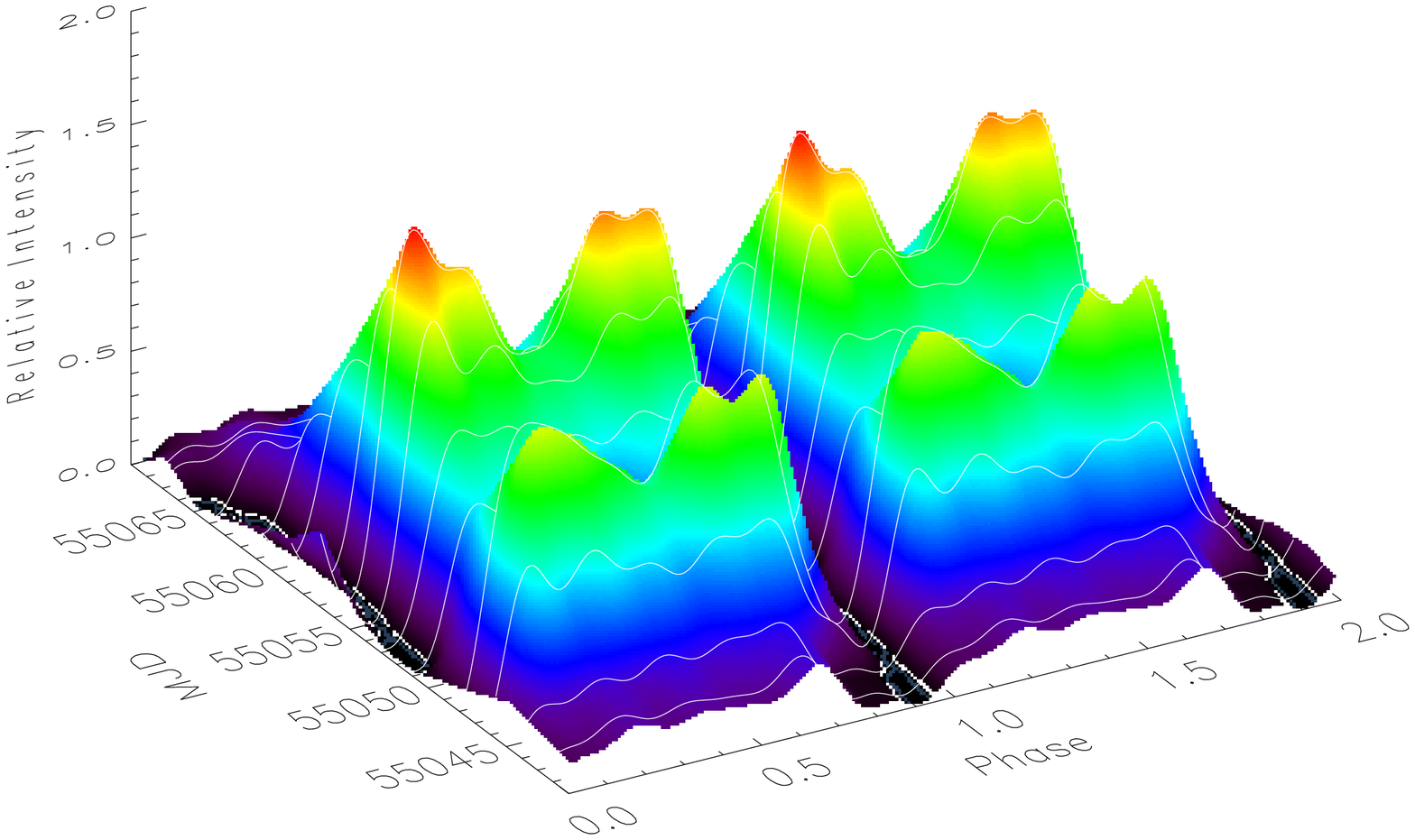}
\caption{Top. 3-D pulse profile evolution with time and flux intensity during the 2009 December  giant outburst  in the 12--25\,keV band. Two phase cycles are shown. Darker colors denote lower flux intensities (see the electronic version for color plots). The two main components of the pulse profile evolve into  a single component (almost flat)  at very low fluxes, as previously seen by BATSE. Bottom. 3-D  pulse profile evolution with time and intensity during the double-peak normal outburst in  2009 November,  in the same band.}\label{prof_giant_lumin}
\end{figure*}

 
\subsection{Pulse Profile Overview}\label{profiles}

During the bright phases of an outburst the pulse profile of A\,0535+26 displays a complex shape in the soft 
X-ray  range, below ~8 keV, and a simple double-peaked profile, with very different spectral shape of the pulses, at higher energies - quite similar to what is  observed in several other accreting pulsars, e.g., SAX\,2103.5+4545 \citep{camero07b}.  A similar overall shape  has been found in a  wide range of observations over several decades in time and outbursts of quite different peak brightness (e.g., \citet{bradt76} and more recently \citet{kretschmar05}, \citet{finger06},  \citet{camero07} and \citet{caballero07}).

Using OSSE data, \citet{maisack96} obtained pulse profiles of  A\,0535+26 in four energy bands (from 35 to 100 keV) during the first and  the second half of the 1994 giant outburst. They saw that the intensity increase  between the first and the second half of the outburst was most pronounced at lower energies, indicating that the spectrum becomes softer as the overall intensity rose. The pulse profile shape was similar to that reported from the previous giant outburst in 1989 with HEXE \citep{kendziorra94,kretschmar96}. 

\citet{bildsten97}  confirmed this behavior based on BATSE data of the 1994 giant outburst. Moreover, they found that at low a  X--ray luminosities ($\le$1.6$\times$10$^{36}$ erg s$^{-1}$) the double-peak pulse profile structure evolved into a broad single sinusoidal-like peak, becoming a top-hat shape at higher energies.


\subsubsection{Recent Giant Outbursts}

Using GBM data  we obtained  for the first time a continuous coverage of the pulse profile evolution for \f\,  during a giant outburst  at  low energies,  in the 8-12 keV band. Fig~\ref{prof_giant} shows the pulse profile evolution with energy for an observation at the peak of the 2009 December giant outburst. The GBM data show a double-peaked pulse profile structure with a large dip between the pulses (two cycles are shown for clarity). The first component spans phases from 0 to $\sim$0.3 and the second one from $\sim$0.5 to 1, with the dip located between phases $\sim$0.3 and $\sim$0.5.  In that observation we can also see how the two main components of the profile evolve in opposite ways with energy. With increasing energy, the profile switches from a weak first and strong second peak to a strong asymmetric primary and almost vanishing secondary peak. In our study we found that in the bright outbursts of 2010 and 2011 the strength of the two peaks is more balanced, albeit at a much lower peak luminosity. During these latter two outbursts the energy dependence of the pulse profile is in line with that observed, e.g., in the 1994 giant outburst by BATSE \citep{bildsten97}, i.e., two peaks of approximately same strength at lower energies with the second decreasing with increasing energy (see Fig.~\ref{prof_giant}).  

In the top panel of  Figure~\ref{prof_giant_lumin} we show a 3-D pulse profile evolution with time and intensity for the 2009 December  giant outburst in the 12-25 keV range.  Darker colors denote lower flux intensities (see the electronic version for color plots). We see that the two main components become more smooth and balanced as the  flux intensity decreases, finally merging into one single component (almost flat)  at very low fluxes, as previously seen by BATSE and/or in normal outbursts.

\subsubsection{Recent Normal Outbursts\label{prof}}

Following the evolution of the pulse profile with luminosity, at energies above 8 keV we found with  GBM a broad single sinusoidal peak at lower luminosities and a clearly double peak profile at high luminosities (see Figure~\ref{prof_giant_lumin}).  With \textit{RXTE}/PCA we found a similar evolution as in previous
studies for this source  by different authors.  With increasing luminosity the low energy profiles became more complex, exhibiting multiple components while the high energy profiles evolved from double to single peaks. \textit{RXTE} data of the 2005 September outburst showed in addition that the pulse profile changed shape around the energy of the fundamental cyclotron line  at 45 keV \citep{camero07}.  Moreover, the pulse shape behavior during the pre-outburst peak in that outburst  was observed to be different than during the main peak (\citet{camero07}, \citet{caballero07}).  This  has not been observed in  other occasions. 

\section{DISCUSSION\label{discuss}}

\subsection{Torque/Luminosity Correlations}

\textit{BATSE} observed Be/X-ray binaries peak spin-up rates of  4.3$\times$10$^{-11}$ Hz s$^{-1}$ (2S 1417--624), 
3.8$\times$10$^{-11}$ Hz s$^{-1}$ (GRO J1750--27) and 8$\times$10$^{-12}$ Hz s$^{-1}$ (4U 0115+634)  \citep{bildsten97}. The discovery outburst of EXO\,2030+375 found it spinning up at a rate of 2.2$\times$10$^{-11}$ Hz s$^{-1}$ \citep{reynolds96}, and  for SAX\,J2103.5+4545 a peak spin-up rate of   1.4$\times$10$^{-12}$ Hz s$^{-1}$ has been reported \citep{camero07b}.  For other type of accreting binaries, e.g. the accreting milisecond pulsars (AMSPs), there are only a few sources of this type where a  spin up could be measured at a rate between 1$\times$10$^{-13}$ and 1$\times$10$^{-12}$ Hz\,s$^{-1}$  \citep[see e.g.][]{papitto08}, compatible with the expected accretion driven spin up. On the other hand, timing noise dominates  the pulse phases observed from other AMSPs, among them SAX\,J1808.4--3658, so that a measure of the spin evolution during  outburst is an extremely difficult task \citep{patruno12, papitto11}. It is to be noted  that the long term spin-down  for this source has remained stable over thirteen 
years with $\dot{\nu}\sim$ -1$\times$10$^{-15}$ Hz\,s$^{-1}$, compatible  with the magnetic dipole spin down of a neutron star with a field of few 10$^8$ G \citep{patruno12}.

Previous studies on \f\, showed the existence of  a strong torque/X--ray luminosity correlation in this BeXRB  system (see e.g. \citet{bildsten97}). Our results using GBM data are in very good agreement with those already published. In  Figure~\ref{long_term_freq} we can see a clear torque-flux correlation, which suggests disk accretion at least in the larger outbursts.  We caution that the apparent frequency rise during the normal outbursts may be an artifact due to the uncertainty of the periastron Epoch. The peak spin-up rate  for the 1994 giant outburst was  11.9$\times$10$^{-12}$ Hz s$^{-1}$ \citep{finger96}, comparable to the 2009 December giant outburst with a  spin-up rate of 10.15(6)$\times$10$^{-12}$ Hz s$^{-1}$.   For the  2011 February giant outburst  the peak of the spin-up rate was  6.36(5)$\times$10$^{-12}$ Hz s$^{-1}$.  This value is smaller than in the previous  giant outbursts, although we note that this  event was weaker. 

Before \textit{BATSE}, it was already known that the spin frequency in some transients decreases between outbursts \citep[and references therein]{bildsten97}.  This was explained  due to the propeller effect, when $\dot{M}$ becomes small enough so that the magnetospheric radius exceeds the corotation radius.
Accretion is then centrifugally inhibited, and material may become temporally attached to magnetic field lines, removing angular momentum and causing the star to spin down \citep{illarionov_sunyaev75}.  \citet{finger94} reported spin-down between outbursts in A0535+26 at a rate of -2.2(6)$\times$10$^{-13}$\,Hz s$^{-1}$.
More recently, before the 2009 December giant outburst, the spin-down rate between outbursts ranged from -1$\times$10$^{-13}$\,Hz s$^{-1}$ to -1.5$\times$10$^{-13}$\,Hz s$^{-1}$. After this event, the spin-down  between the last normal outburst and the last giant outburst occurred in 2011 February  happened at an approximate rate  of -3.3$\times$10$^{-13}$\,Hz s$^{-1}$.

 \f\, also exhibits global spin down trends during  long periods of quiescence.  Assuming that in quiescence \f\, enters the subsonic propeller regimen, it is expected to spin-down at a rate of  $\dot\nu=-4\pi\nu^2\mu^2\,(GM)^{-1}I^{-1}$   $\sim$\,-2.4$\times$10$^{-14}$ Hz s$^{-1}$, where  $\mu$ is the neutron star magnetic moment, $M$ the mass (1.4 M$\odot$) and $I$ the moment of inertia\citep{henrichs93}. Based on our observations, during  the quiescence period between 2005 and 2008, \f\, showed a spin down  trend  with  $\dot\nu$\, $\sim$\,-1.01(5)$\times$10$^{-13}$ Hz s$^{-1}$.  These two values are not in agreement and therefore  this behavior cannot be well explained. On the other hand, mass prevented from accreting by the presence of a centrifugal barrier does not necessarily have to leave the system  \citep[and references therein]{d'angelo12}. These authors  found that the accretion from a disk on to a magnetosphere tends to take place in 'trapped' states, where the accretion  can be steady or cyclic. Two forms of cyclic accretion are found in this study,  a) the accretion--dead disk cycle  \citep{sunyaev_shakura77} and b)  where a pile-up of mass at the centrifugal barrier outside corotation is followed by an episode of accretion emptying the pile and the beginning of a new cycle.  In this model, the effect of these cycles is to increase the average spin-down torque on the star.  However, whether this model may be applicable to \f\, has not been explored.

\subsection{Quasi Periodic Oscillations}

\citet{reig08} carried out a rapid spectral and timing variability study of Be/X-ray binaries
during type II outbursts. In this study, the aperiodic variability of the four Be/X-ray binaries was described with a relatively low number of broad Lorentzian components, and with peaked noise (for some of the sources as QPO noise)  between  the low-frequency and high frequency noise. In addition, the  flat-topped noise found at  lower frequencies for sources in the horizontal branch of the HR--Intensity diagram, turned into power-law noise in the diagonal branch. Some of these components can be associated with the same type of noise as seen in low-mass X-ray binaries, although the characteristic frequencies are about one order of magnitude lower \citep[and references therein]{reig08}. 

The origin and properties of QPOs are poorly understood. In accretion-powered X-ray pulsars, QPOs fall in the milliHz range. The frequencies of these QPOs range from 1–-400 mHz, a few orders of magnitude lower than in  low-mass X-ray binaries \citep[and references therein]{reig08}.  It is generally accepted that QPO features correlated  with flux/spin-up are a powerful indication of the presence of an accretion disk.   According to the beat frequency (BF) model \citep{Alpar85, Lamb85}, blobs of matter orbiting at the inner disk boundary at approximately the Keplerian frequency $\nu_{k}$ , 
are  gradually removed through the interaction with the neutron star magnetosphere.  The magnetic field rotates with the neutron star at a frequency $\nu_{s}$, modulating the accretion rate and source luminosity. Therefore the QPO frequency  results from the beat between the orbital frequency of the blobs at the inner disk boundary  and the spin frequency  $\nu_{s}$
of the neutron star ($\nu_{QPO}$ = $\nu_{k}$ -- $\nu_{s}$).  In the Keplerian frequency (KF) model \citep{vanderKlis87}, the inner edge of the accretion disk contains structures that persist for a few cycles around the neutron star, and modulate the observed flux by obscuration producing a QPO at the Keplerian frequency $\nu_{k}$.   At lower mass accretion rates torques caused by the interaction of the magnetic field with the accretion disk also need to be considered. In the magnetically threaded disk model  \citep{ghoshlamb79},  a magnetic torque is generated by the neutron star magnetic field threading the disk.  In their model, the coupling between the neutron star and the accretion disk occurs in a broad transition zone located between the flow inside the neutron star magnetosphere and the unperturbed disk flow.  Another  model, the magnetic disk precession model \citep{shirakawa_Lai02}, attributes the mHz QPO in X-ray pulsars to warping/precession modes induced by magnetic torques near the inner edge of the accretion disk. We note that this model has only been applied to weakly magnetized neutron stars in low-mass X-ray binaries.

For A\,0535+26, \cite{finger96} detected for the first time a  mHz QPO  during the 1994 giant outburst observed by BATSE.  This QPO was detected for 33 days during the outburst. During this time the center frequency of the QPO rose from 27 to 72 mHz and then fell again to 25 mHz.  These observations showed a strong correlation between  the X--ray flux and  spin-up rate for A\,0535+26,  reinforcing the  idea of  the presence of an accretion disk \citep[and references therein]{finger96}.   These authors successfully applied both the Beat Frequency and the Keplerian models to these observations.   These results showed that the QPO center frequency was, like the spin-up rate and the flux, controlled principally by the mass accretion rate and linked  with the inner region of the accretion disk. In addition, \cite{bozzo09} discussed the physical equations that determine the position of the inner disk boundary by using different prescriptions for the  neutron star-accretion disk interaction.  They applied the results to several accreting pulsars showing both QPOs in their X--ray flux  and  torque reversals.  They also concluded that  for A\,0535+262  a good agreement between the BF model and the GL model  was obtained. 

In the present work,  we reported on the detection of a hard 30--70 mHz QPO using GBM data for  A\,0535+26,  during the giant outburst  occurred in 2009 December \citep{fingerwilsoncamero2009}. The QPO was detected for 24 days and the center  frequency rose from 30 to 70 mHz and went back to 30 mHz. Once again, as in the QPO detected by BATSE, a strong QPO center frequency/X--ray flux correlation was found.  The flux QPO frequency correlation may be explained with the BF  model. However, the fact that the QPO was not detected by GBM in the most sensitive  band of 12--25\,keV, but only above this one, is inconsistent with this interpretation. It is out of the scope of this work to elaborate a sophisticated theory reproducing these observations, however we will speculate with  the origin of this inconsistency. In the KF model, the QPO was explained as produced  by quasi-periodic oscillations in the thick inner accretion disk which partially obscures the radiation emitted in our line of sight.  We can identify the major contribution of the hard X-ray component  in the spectrum of \f ~ with emission from the lower part of the accretion column and from the central disk regions,  and  the low energy component with emission from scattering from the neutron star surface.  One possible explanation for our observations is that the regions where  hard X-rays originate might be obscured in our direction, while for the  softer X--ray radiation  no QPO is produced because  any appreciable emitting region is occulted. 

QPOs studies may also allow us to constrain the accretion flow. Considering the BF model and the relationship between the $\nu_{QPO}$ and the Keplerian radius ($r_{\rm k} =  (GM/4\pi^2 \nu_{k}^2)^{1/3}$),  we obtained  a value for $r_{\rm k}$ of $\sim$9\,100 km ($\sim$9\,960 km using the KF model),  for an observation at the peak of the 2009 December giant outburst.  For \f\,the last detection of a QPO by GBM was in 2009 December 27, with a  $\nu_{QPO}$ of $\sim$33.771 mHz. This means a   $r_{\rm k}\sim$13\,500  km (with KF model $\sim$16\,060 km),  consistent with the standard picture of an accretion disk moving outwards as the accretion rate decreases.

\citet{finger96} concluded that the QPO in \f\, is consistent with the beat frequency model, with the QPO occurring at the beat frequency  $\nu_{QPO} = \nu_k - \nu_{ns}$. They however believed that the fundamental cyclotron line energy was near 110 keV. This is now known to be the first harmonic, with the fundamental line near 50 keV \citep{kendziorra94}.  We therefore will reconsider this interpretation.

This model predicts two relationships between observed quantities \citep{finger96}. 
The first is between the pulsar spin-up rate $\dot{\nu}_{ns}$ and the Keplerian frequency $\nu_k$,
\begin{equation}
   \dot{\nu}_{ns} = \aleph_N \nu_k^2~~{\rm with}~~
   \aleph_N = 1.6K^{7/2} B^2 R^6 (GM)^{-1} I^{-1}~, \label{aleph_n}
\end{equation}
where $B$ is the surface magnetic field at the magnetic pole, 
$R$ the neutron star radius, ${G}$ Newton's gravitational constant, 
$M$ the neutron star mass, and $I$ it's moment of inertia. 
$K$ is the ratio of the inner disk radius $r_0$ to $\mu^{4/7} (GM)^{-1/7} \dot{M}^{-2/7}$, which is expected to be near 1 (in particular $K\approx$\,0.91 gives the Alfv\'{e}n radius for spherical accretion).
Here $\mu = 0.5 B R^3$ is the magnetic moment, and $\dot{M}$ the mass accretion rate. 
The value determined from the 1994 giant outburst was 
$\aleph_N = (1.91 \pm 0.02) \times 10^{-9}~{\rm erg~cm}^{-2} {\rm s}^{4/3}$.

The second relationship is between the flux $F$ and the Keplerian frequency $\nu_k$,
\begin{equation}
   F = \aleph_F \nu_k^{7/3}~~{\rm with}~~
   \aleph_F = 1.4\alpha\beta d^{-2}K^{7/2} B^2 R^5 (GM)^{-2/3}~, \label{aleph_f}
\end{equation}
where $\alpha$ is the fraction of the bolometric luminosity in the observed energy band,
$\beta$ accounts for beaming, and $d$ is the source distance. From the peak QPO frequency
of $69 \pm 1$ mHz in the 2009 giant outburst, and the 3--50 keV peak flux of 
$1.25 \times 10^{-7}~{\rm erg~cm}^{-2}~{\rm s}^{-1}$
we find $\aleph_F = (4.7 \pm 0.2) \times 10^{-5}~{\rm erg~cm}^{-2} {\rm s}^{4/3}$.

If we determine $B = 4.3 \times 10^{12}$\,G from the cyclotron line energy of 50 keV, 
and adopt the nominal values $\alpha = \beta = 1$, $d = 2$  kpc, $R = 10^6$\,cm, and 
$M = 1.4 M_\odot$, then from equation \ref{aleph_f} and the measured value of  
$\aleph_F$ we would conclude that $K \simeq 2.4$. This would imply that the
QPO is not formed at the inner region of the accretion disk but at a radius much 
further out, or that the flow at the inner region of the disk is sub-Keplerian. We point out that the resulted discrepancy of the value of $K$ between the 1994 ($K\sim$ 1) and 2009 giant outbursts is due to  different assumptions, in particular to that of the implied magnetic field and the assumed $\alpha$, $\beta$ and $d$. In \citet{finger96} they proposed values for $\alpha\approx 0.5$, 2 $\leq \beta \leq$ 3, 1.3 kpc $\leq d \leq$ 2.6 kpc and, as stated before, with the fundamental of the  cyclotron line energy at  110 keV.

\begin{figure}[h]
\includegraphics[width=9cm,height=7.5cm]{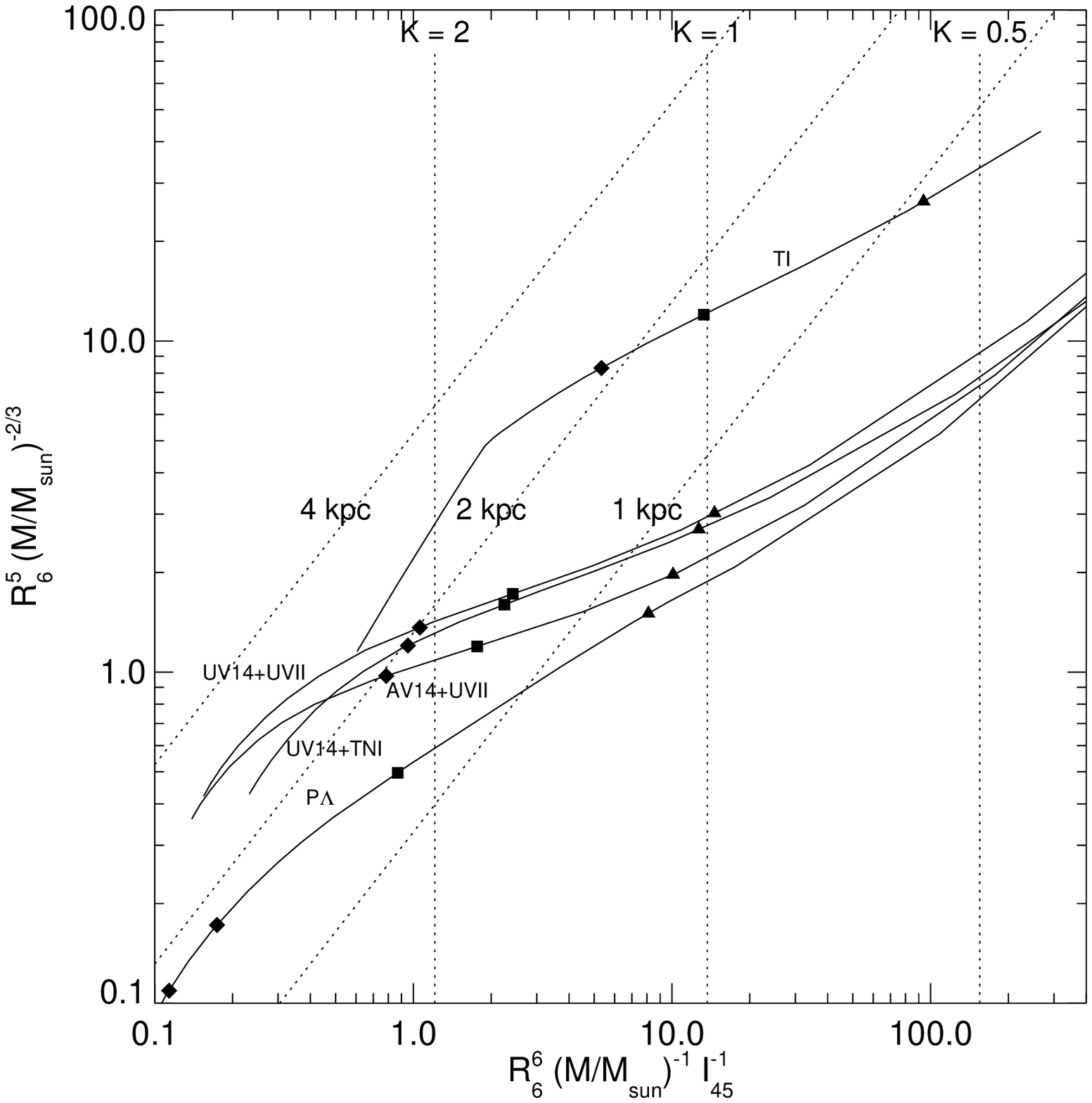}
\caption{The combination of neutron star parameters in equations \ref{aleph_n}
and \ref{aleph_f} are  plotted for the set of neutron star models tabulated in
\citet{wiringa88}.  The TI curve is for
the tensor interaction model of \citet{pandharipande75} which is a very
stiff equation of state.  The P$\Lambda$ curve is for the \citet{pandharipande71} 
hyperon model,  which is a very soft equation of state. The AV14+UVII,
UV14+UVII and UV14+TNI models are from \citet{wiringa88}.  The 
symbols on the curves give the points where the mass is
0.5$M_\odot$ (triangles), 1.0$M_\odot$ (squares), and 1.4$M_\odot$ (diamonds).
The vertical dotted lines are for constant values of $K$, which determines
the ratio of the inner disk radius to the Alfv\'{e}n radius ($K\approx$\,0.91 gives the Alfv\'{e}n radius for spherical accretion). The sloped dotted lines are for constant values of $d(\alpha\beta)^{-1/2}$, where $d$ is
the source distance,  $\alpha$ the bolometric correction factor, and $\beta$ the
beaming factor.\label{ns_plot}}
\end{figure}

The other possibility is that the adopted values for the parameters
needed to convert the measured flux to the true luminosity, and the neutron star parameters,
can be adjusted in a plausible way so that the beat frequency model agrees with
observations. In Fig. \ref{ns_plot} we show the combinations of neutron star
parameters in equations \ref{aleph_n} and \ref{aleph_f} for a set of neutron star models. The plot also shows lines
of constant $K$ and $d(\alpha\beta)^{-1/2}$ assuming a magnetic field of 
$B = 4.3 \times 10^{12}$\,G the measured values of $\aleph_N$ and $\aleph_F$. 
The conversion of the measured flux to the true luminosity depends on $d(\alpha\beta)^{-1/2}$.
The most recent measurement of the source distance $d$ is $\sim 2$\,kpc \citep{steele98},
with previous measurements ranging from 1.3 to 2.6 kpc. The bolometric fraction $\alpha$ 
probably above 80$\%$. The beaming factor $\beta$ result from the fact that the
phase averaged flux only samples the pulsar's spherical beam pattern over a cone. This term
may vary from 1 with viewing direction by $\sim$50\%. The factor $K$ is expected to be
in the range of about 0.5-1.0. So overall we think the beat frequency model is 
plausible for neutron star parameters in the region with $K$ in the range 0.5-1, and  
$d(\alpha\beta)^{-1/2}$ in the range 1-3 kpc. This contains neutron stars with stiff equations of state, and masses of 1 $M_\odot$ or less. We note, however, that the current mass estimates for neutron stars are in excess of 1.2 M$_{\odot}$, with the majority of normal pulsars lying in the mass range 1.4--1.8 M$_{\odot}$ (\citet{ozel12}, \citet{steiner10}).

Reynolds et al. (2010) found broadened Fe ${\rm K} _\alpha$ profiles in two Chandra grating observations that occurred during the tail of the 2009 giant outbursts. They fit the lines with Gaussians with sigmas of $\simeq$ 5100 and 4\,100 km s$^{-1}$ respectively. They interpreted broadening as due to the azimuthal velocity at the inner edge of the accretion disk. These observations occurred at approximately 50$\%$ and 40$\%$ of the peak flux. If beat frequency model for the QPO is correct, we can predict the Keplerian frequency at the disk inner edge, which scales as the $F^{3/7}$, giving $\nu_k =~$ 60 and 53 mHz respectively for the two Chandra observations. The azimuthal component of velocity at the inner edge is then $V_\phi = (2\pi \nu_k GM)^{1/3}$ or 3700 and 3500 km s$^{-1}$. Given that these set the maximum amplitude of the projected velocity distribution, and these will be decrease further by the (unknown) inclination of the disk from the line of sight, we find this prediction  in disagreement with the observations.

\subsection{Pulse Profile Study \label{disccus_prof}}

Our study mostly is in line with previous observations (see Sect.~\ref{profiles}). At medium to high source luminosity we find a multi-peak pulse shape at energies below $\sim$8 keV, evolving into a double-peak structure at higher energies. The high-energy pulse shape evolution with luminosity also agrees with previous works, in that the profiles evolve towards a single broad peak as luminosity decreases  (see Fig.~\ref{prof_giant}).

The giant outburst of 2009, the brightest in our sample of outbursts monitored by GBM, is peculiar in that the two main components of the pulse profile evolved in opposite ways with energy. To our knowledge, this has been never reported before. In contrast, for the other bright outbursts in 2010 and 2011, our results are in good agreement with the  energy dependence seen in the 1994 giant outburst \citep{bildsten97}. At the moment we have no explanation for the differences between the outbursts.  The mechanisms leading to different kinds of outbursts are not well understood for different sources and  currently there is no good model to describe  A\, 0535+26. It remains unknown whether  the giant outburst detected by  BATSE in 1994 was in fact different from the giant outburst detected by GBM in 2009.  With BATSE  the pulse profile study was carried out  above 20\,keV, preventing us from performing any comparison. Above 20\,keV  the  evolution of the pulse profile is consistent with GBM.

Interpreting pulse profiles in terms of the underlying emission pattern is far from evident since various effects, including highly anisotropic cross-sections for emission and scattering and light-bending around the neutron star must be taken into account \citep{kraus03}.  For A\,0535+26 \citet{caballero11a} have disentangled the contribution of the two single pole components for  A\,0535+26 using pulse profiles from the 2005 August normal outburst. The beam pattern was interpreted in terms of a geometrical model that consists of a hollow column plus a halo of scattered radiation around the neutron star surface. In the model by \citet{ibragimov09}, the presence  (or not) of a secondary maximum in the pulse profile is interpreted by the appearance (or disappearance) of a secondary antipodal spot. This model explains that obscuration of the antipodal spot  by  variations of the inner disk radius reproduces the variety of the pulse profiles observed in the accreting millisecond pulsar SAX J1808.4-3658.  The applicability to \f\, is not yet certain.

An intriguing case was the pulse shape behavior at the rise of the  2005 September normal outburst \citep{camero07,caballero07}.   One of the components of the multi-peak structure at low energies became more prominent compared with the pulse profile during the main outburst. The evolution with energy was also different, evolving into one strong simple peak as the energy increased.  A  different accretion regime was suggested during the pre-outburst phase than during the main outburst. A possible explanation  was bursts of wind-like accretion from capture of low angular momentum material, while the higher angular momentum material forms an accretion disk, which feed the main part of the outburst.  \citet{caballero07,caballero08} found a change of the cyclotron line energy up to  $\sim$50 keV during this period and no significant change  in the spin frequency up to the time the main outburst began. \citet{postnov08} proposed  that magnetospheric instabilities developing at the onset of the accretion were the origin of the flaring activity.  This may explain the difference in the cyclotron line energy and energy dependent pulse profiles, that remained stable as the neutron star starts spinning-up and the flaring activity disappears.  We point out that for \f  while the profile changes in the later stages of an outburst  the line actually seems stable \citep{caballero10}.

In contrast, during the double-peaked outbursts of 2009 August  and 2010 July (see Figure~\ref{2xnormal})  no changes in the shape of the pulse profile were found  in the present work.  Also no change of the cyclotron line energy was found by \citet{caballero09} for the former outburst. We leave as an open question  whether these pre-outbursts peaks are related to the formation and/or evolution  of the disk around the Be star.

\subsection{Be-disk Neutron Star Interaction}

In this study we have shown how the H$_\alpha$ line is practically always in emission, except for a small period around MJD 51000, telling us about the presence of a persistent but variable Be-disk.  In the long-term, the  optical brightening of the Be star, besides large negative values of the H$_\alpha$ EW, seems to precede the apparition of episodes of high X--ray luminosity (and neutron star spin-up), including giant X--ray outbursts, indicating that the variations in the circumstellar structure are the precursors of this X--ray activity. This behavior has been observed in other BeXRBs, as for instance in EXO\,2030+375 \citep{wilson08}. We have also observed that the onset of both normal and giant outbursts are initiated near periastron. Although this does not fit the \citet{stella86} picture of type II outbursts with timing unrelated to the underlying orbital period.   

\begin{figure}[!]
\hspace{-0.5cm}
\includegraphics[width=9.25cm,height=9.75cm]{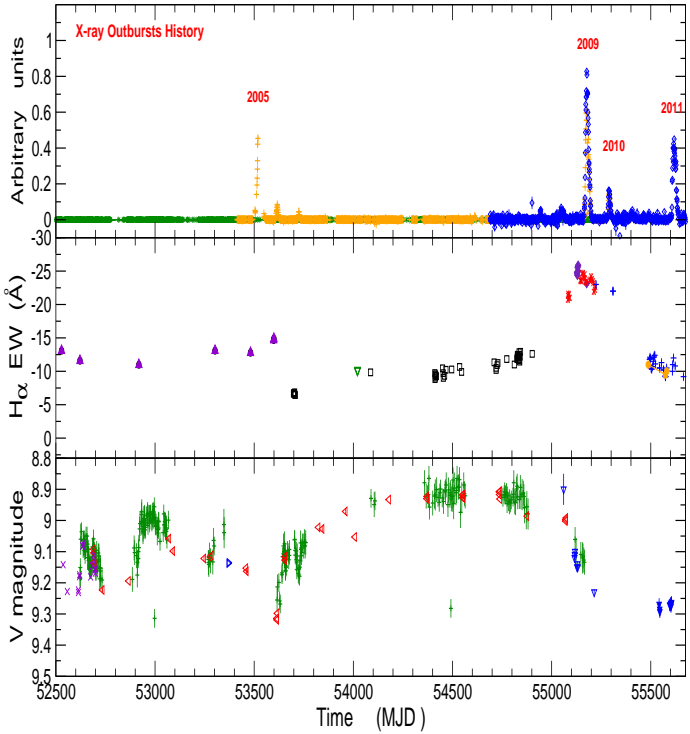}
\caption{Zoom of Fig~\ref{long_term_flux}  from $\sim$2002 August  up to 2011 March.  We can see the anticorrelation of the H$_\alpha$  EW and the visual magnitude, with each giant X-ray outburst occurring during a decline phase of the optical brightness.\label{zoom_2}}
\end{figure}

Taking a closer look, the Be star lost and renewed its circumstellar disk during the period from $\sim$1996 until 2005. After this period the H$_\alpha$ EW and the optical brightening were no longer correlated. Fig~\ref{zoom_2} shows that after $\sim$MJD 52500 (2002 August) each giant X--ray outburst occurred during a decline phase of the optical brightness. It is interesting that the H$_\alpha$ showed a strong emission, while its optical brightness had a decreasing trend before and after the 2009 giant X--ray outburst. That is, the H$_\alpha$ EW and the V-band brightness showed an anti-correlation.  \citet{yan12} propose that this anti-correlation may indicate that a mass ejection event took place before the 2009 giant X--ray outburst.

The H$_\alpha$ EW measurements (see e.g. \cite{Haigh04}, \citet{grundstrom_gies06}) implies this was the largest Be-disk observed in the system since its discovery.  This may explain the unusually active period of A\,0535+26 in X--rays. The decreasing trend and drastic optical spectroscopic variability after this  2009 December giant outburst has also been recently reported by \citet{moritani11}.  However, their measurements of the  H$_{\alpha}$  EW  before and during the 2009 giant outburst are much lower than our values and those from \citet{yan12}. Most likely these differences are not real, but the product of two different instrument systems (spectral resolution $\&$ continuum determinations).


\vspace*{0.25cm}

\subsubsection{Recent $H\alpha$ Line  Profile Evolution}

It is interesting to compare (Fig~\ref{halfa_prof}) the H$\alpha$ line profiles of the disks before, during, and after the giant outbursts in 2009 December (left hand panel) and 2011 February (right hand panel), as observed by the FRODOSpec spectrograph ($R\sim$\,5000) on the Liverpool Telescope. This data presents a far higher time resolution ($\sim$ few days) than that typically obtained around the time of previous bursts ($\sim$\,1 month).

Throughout the 2009 outburst the H$\alpha$ profile was consistently single peaked with a slightly asymmetric profile characterized by a broader blue wing, and no significant change can be seen around the time of outburst.  A comparison with the higher resolution ($R\sim\,30,000- 60,000$) but more infrequent spectra presented by \citet{moritani11}, over the same time period, shows the same basic structure, although they are able to resolve the peak which shows a  double peak with velocity separation of $\sim$\,100 km\,$s^{-1}$.   On occasions small additional absorption and emission features are also seen in their spectra. We note that \citet{naik11} did not find any major changes in a series of JHK spectra taken during the X-ray quiescent phase and the giant X-ray outburst in 2011 February–March.

In late 2010 (prior to the 2011 February outburst) the disk can be seen to be in a very different state, with an H$\alpha$ equivalent width of only half that seen in late 2009 and early 2010.  The profile is double peaked with a separation corresponding to $\sim$\,270 km\,$s^{-1}$.  Most striking are the rapid V/R (violet/red) variations in the strength of the two peaks.  These have a period of $\sim$\,25 days and can be seen to repeat over two cycle from late 2010 October  until mid 2010 December.   This is far shorter than the typical $\sim$ 1000 day periods seen in isolated Be stars \citep{hummel98}.  V/R variations with a timescale of $\sim$\,1 year were noted from \f in the period following the giant outburst of 1994 Feb/Mar  by \citet{clark98} (from 1994 September until late 2006),  who attributed them to the presence of a global one-armed oscillation \citep{okazaki91}. From mid 2010 December the profile changes again, with the blue peak remaining dominant and the red wing broadening as a red component moves to higher velocities.  No specific change at the time of the  2011 February outburst is obvious, with the red wing simply continuing its move to higher velocities. 

A general pattern of behavior may start to be inferred from the behavior of the system following the 1994 March  and 2009 December giant outbursts, where the disk becomes weaker and shows $V/R$ variability commencing $\sim$\,6 months following a giant outburst.  Following the model of \citet{okazakinegueruela01} (which is consistent with the  variability of the circumstellar disk after the 2009 giant X-ray outburst), it is important to note that those disk changes are not caused by the outburst, but rather that the outburst is a signal that changes in the disk are underway, presumably due to changes in the truncation resonance radius.  If this pattern is to be believed, then when the source becomes observable again in 2011 September, we would expect the disk strength and associated line profile to differ from that presented in 2011 April.  In this overall picture what is still an open question is the  mechanism/s that  trigger a giant outburst.

\section{Conclusions}

\f\, exhibits an unpredictable X--ray behavior, with long periods of quiescence related to an absence of H$_\alpha$ in emission, with periodic normal outbursts generally taking place before and/or after a giant outburst.  Some peculiar normal outburts of \f present a double peaked structure, with the origin of these pre-outburst peaks still unknown.  Our study of the spectral variability of this source using GBM data from recent outbursts shows that there is a hardening of the spectrum as the flux intensity increases.

Our pulse profile study for \f\, confirms in general previous results. However, for the brightest outburst monitored by GBM in 2009 December, we found that the two main components of the pulse profile evolved in opposite ways with energy, which  to our knowledge  has been never reported before.  A hard 30-70 mHz  X-ray QPO was detected  with GBM during this  giant outburst. The new insights are that it becomes stronger with increasing energy and disappears at energies below 25\,keV.  Current models cannot explain these observations. 

In spite of the fact that in the long-term a strong optical/X--ray correlation has been found for this BeXRB, however it is still difficult to foresee when a giant outburst is going to take place and therefore what the trigger mechanism might be.   On the other hand, in the medium-term  the H$_\alpha$ EW and the V-band brightness show an anti-correlation after $\sim$2002 August (MJD 52500).  Each giant X-ray outburst occurred during a decline phase of the optical brightness, while  the H$_\alpha$ showed a strong emission. We would like to point out  that the large event that came after the 2009 December giant outburst does not follow the standard classification as either Type-I or Type-II outburst.  

The Be circumstellar disk was exceptionally large just before the 2009 December giant outburst. Most likely this was the origin of the unusual recent X--ray activity of this source.  In late 2010  and prior to the  2011 February outburst,  rapid V/R variations are observed in the strength of the two peaks of the H$_\alpha$ line, with a period of $\sim$\,25 days, suggesting  the presence of a global one-armed oscillation. A general pattern of behavior is proposed in this work,  where the disk becomes weaker and shows V/R variability beginning $\sim$\,6 months following a giant outburst.

\acknowledgments {Acknowledgments. A.C.A. and M.H.F. acknowledge support from NASA grants\linebreak NNX08AW06G and NNX11AE24G. P.J. acknowledges support from NASA Postdoctoral Program at NASA Marshall Space Flight Center, administered by Oak Ridge Associated Universities through a contract with NASA.  J.G.S. acknowledge the  Instituto de Astrof\'{i}sica de Andaluc\'{i}a (CSIC) for allowing us to access to the  0.9 and 1.5m telescopes at the OSN at the Observatorio de Sierra Nevada (Spain).  A.C.A thanks to Alessandro Pappito for providing very useful information on milisecond accreting X-ray pulsars. We  thank to all the \textit{Fermi}/GBM Occultation team for its help and in general to all the GBM team based in Huntsville, Al.}


\end{document}